\documentclass[12pt,preprint]{aastex}

\newcommand{\eg}{{\it e.g.}\ }
\newcommand{\ie}{{\it i.e.}\ }

\begin{document}


\title{The Thousand Asteroid Light Curve Survey\altaffilmark{1}}
\author{Joseph Masiero\altaffilmark{2}, Robert
  Jedicke\altaffilmark{2}, Josef \v{D}urech\altaffilmark{3}, Stephen
  Gwyn\altaffilmark{4}, \nl Larry Denneau\altaffilmark{2}, Jeff Larsen\altaffilmark{5}}

\altaffiltext{1}{Based on observations obtained with MegaPrime/MegaCam, a joint project of CFHT and CEA/DAPNIA, at the Canada-France-Hawaii Telescope (CFHT) which is operated by the National Research Council (NRC) of Canada, the Institute National des Sciences de l'Univers of the Centre National de la Recherche Scientifique of France, and the University of Hawaii.}
\altaffiltext{2}{Institute for Astronomy, University of Hawaii, 2680 Woodlawn Dr, Honolulu, HI 96822, {\it masiero, jedicke, denneau@ifa.hawaii.edu}}
\altaffiltext{3}{Astronomical Institute, Faculty of Mathematics and Physics, Charles University in Prague, V Holesovickach 2, 18000 Prague, Czech Republic, {\it durech@sirrah.troja.mff.cuni.cz}}
\altaffiltext{4}{Canadian Astronomical Data Centre, Herzberg Institute of Astrophysics, Victoria, BC, V9E 2E7, Canada {\it Stephen.Gwyn@nrc-cnrc.gc.ca}}
\altaffiltext{5}{Physics Department, United States Naval Academy, Annapolis, MD 21402, {\it larsen@usna.edu}}

\begin{abstract}
We present the results of our Thousand Asteroid Light Curve Survey (TALCS)
conducted with the Canada-France-Hawaii Telescope in September 2006.  Our
untargeted survey detected $828$ Main Belt asteroids to a limiting magnitude
of $g'\sim22.5$ corresponding to a diameter range of $0.4~$km $\le D \le
10~$km.  Of these, $278$ objects had photometry of sufficient quality to
perform rotation period fits.  We debiased the observations and light curve
fitting process to determine the true distribution of rotation periods and
light curve amplitudes of Main Belt asteroids.  We confirm a previously
reported excess in the fraction of fast rotators but find a much larger
excess of slow rotating asteroids ($\sim15\%$ of our sample).  A few percent
of objects in the TALCS size range have large light curve amplitudes of $\sim
1$~mag.  Fits to the debiased distribution of light curve amplitudes indicate
that the distribution of triaxial ellipsoid asteroid shapes is proportional
to the square of the axis-ratio, $(b/a)^2$, and may be bi-modal.  Finally, we
find six objects with rotation periods that may be less than $2~$hours with
diameters between $400~$m and $1.5~$km, well above the break-up limit for a
gravitationally-bound aggregate.  Our debiased data indicate that this
population represents $<4\%$ of the Main Belt in the $1-10~$km size range.

\end{abstract}


\section{Introduction}
\label{intro}
Following the call by \citet{pravecAIII} for an asteroid rotation survey free
from biases against low amplitude and long period objects, we have conducted
a large, untargeted survey of small Main Belt asteroids.  Our Thousand
Asteroid Light Curve Survey (TALCS) was designed to use the flexibility of
the Canada-France-Hawaii Telescope (CFHT) queue schedule observing to
discover and obtain light curves for a large and random sampling of Main Belt
asteroids with controlled and understood biases.

Rotation measurements of asteroids are one of the primary ways of deriving
physical properties of these bodies from Earth-based instruments.  If assumed
to have shapes described by triaxial ellipsoids and constant albedos across
their surface, the intensity of the reflected light can be well described by
a simple sinusoidal relationship \citep{pedro03}.  More complicated models
are needed to describe the light curves generated by asteroids with realistic
shapes (see Fig.~1 from \citet{sullivanAIII} for a composite of asteroid
photos from {\it in situ} studies).  \citet{harris89} presented a method for
describing complex light curves with multi-order sinusoids which has become
the standard for fitting light curves with a high-density of data
\citep[\eg][]{pravec00,pravecAIII}.  For asteroids with sparse data
sampling, light curve inversion has become a powerful tool for finding
rotation rates while its application to high density data has provided
excellent models of asteroid shapes
\citep{mikko04,mikko07,durech07,durech09}.

\citet{pravecAIII} discuss the relationship between asteroid spin rate and
diameter using all published periods for Main Belt and Near Earth asteroids.
They point out a strong barrier in rotation period at $P\approx2~$hours for
objects larger than $D>150~$m.  Gravitationally bound aggregates of smaller
boulders will have a spin rate limit of $2-3~$hours depending on the density
of the composite rocks and their characteristic size \citep{pravec00}.
Objects with diameters smaller than $150~$m have been observed rotating
considerably faster than this limit and can be explained as monolithic rocks
with internal tensile strength.  The two hour spin limit implies that bodies
larger than $150~$m are unable to survive as monoliths in the collisionally
processed inner Main Belt \citep{harris96} which also feeds the Near Earth
Object (NEO) population \citep{bottkeAIII}.

The spin rates of a collisionally evolved system should fit a Maxwellian
distribution as the spin vectors parallel to the plane of motion should be
normally distributed around zero while the orthogonal vector will have a
slight asymmetry due to the bulk motion of the system \citep{salo87}.  This
produces a distribution of the form:
\[N\propto~\frac{\omega^2}{a^3}~e^{(-0.5~\omega^2~a^{-2})}\]
where $\omega$ is the spin rate and $a$ is a constant, which does an
excellent job of describing the spin rates observed for large ($D>40~$km)
Main Belt asteroids \citep{pravecAIII}.  This relation breaks down at smaller
sizes \citep{pravecAIII,polishook08} when non-collisional forces, in
particular the Yarkovsky-O'Keefe-Radzievskii-Paddack (YORP) effect
\citep{rubincam00}, create excesses of objects with very slow and very fast
rotation rates \citep{vok02}.

\citet{holsapple07} proposed a size-dependent strength for asteroids that can
explain both the cutoff in rotation rates of the largest objects at
$P\sim2~$hours as well as the existence of rapidly rotating small NEOs
without requiring a sharp transition in composition or evolution.  It would
also imply that there should exist some objects rotating faster than the
$2~$hour ``spin limit'' in the $0.15-5~$km size regime.  \citet{pravecAIII}
discuss a single object found to be above the rotation limit (2001 OE$_{84}$,
a Mars Crossing asteroid with period of $\sim29.19$min and size of
$\sim0.9~$km) however the authors indicate that this object is likely an
exceptional one.  Recently, \citet{pravec08} performed a similar study to
ours of small inner-Main Belt asteroids.  They find an excess of slow
rotators in their $3-15~$km diameter range sample but a flat rotation rate
distribution from spin rates of $1-9.5~$rev day$^{-1}$.  They explain this
distinctly non-Maxwellian distribution by modeling the YORP effect on small
bodies and showing that when the YORP timescale is dominant any previous
distribution is erased.

\section{Observations}

The TALCS program was designed to obtain orbits, colors, and basic light
curves for approximately $1000$ Main Belt asteroids.  Using the
Canada-France-Hawaii Telescope's (CFHT) MegaCam instrument \citep{megacam} we
surveyed approximately $12~$deg$^2$ of the ecliptic (\ie $12$ MegaCam
pointings of $\sim1~$deg$^2$ each) for six nights across a two-week period in
September 2006.  Using the flexibility that is only possible under
Queue-scheduled observing, we spread our allocated time over the instrument's
run, observing only 3-4 hours per night.  The asymmetric spacing increased
our sensitivity to slower rotating objects while giving us a longer orbital
arc for better orbit determination than would be possible for the same amount
of observing performed in a classical mode.

The survey was centered $1~$hr away from opposition to minimize the
opposition effect, a non-linear increase in the reflected light at low phase
angles ($\alpha \le 5^\circ$).  Though this would have increased our
sensitivity to fainter targets it would have complicated the light curve
analysis.  The phase angles of our targets ranged from $4^\circ-13^\circ$
depending on their distance from the Sun and the times of observation.  To
reduce loss of objects between nights the center of our survey pattern
drifted each night to follow a hypothetical mid-Main Belt asteroid on a
circular orbit.  Table~\ref{tab.survey} provides the dates, central position
and setup for each night of observing.  One night (20 Sep 2006) was devoted
to a subset of the whole field to decrease the time between exposures and
thus improve light curve resolution.  Additionally, all images on the night
of 17 Sep 2006 were taken with the $r'$ filter (as opposed to $g'$ for all
other nights) to allow for color determination of the asteroids.

Over the two-week observing run we obtained $1079$ images totaling
$20.4~$hours of queue time.  Initial data reduction was accomplished on
location by the standard Elixir pipeline \citep{elixir}.  Post-processing was
performed using the MegaPipe service provided by the CADC \citep{megapipe}.
Astrometric calibrations were initially based on the USNO catalog with a
deeper internal catalog generated from all stationary objects in the survey.
Photometric calibrations were based on the initial Elixir calibrations and
then finalized based on a set of secondary standards from the photometric
nights.  Images taken on non-photometric nights were calibrated with these
secondary standards.  The systematic astrometric error was $\sim0.05''$ and
systematic error on relative and absolute photometry was $\sim0.015$ and
$\sim0.03~$mag respectively.

Source identification used SExtractor \citep{sex} with the requirement of a
detection $\ge1.5~\sigma_{sky}$ on four contiguous pixels.  Pre-filtering of
the images yielded an effective signal-to-noise limit of $4~\sigma$ above
background.  Source lists were then cleaned based on the half-light radius of
the detection, selecting point-like PSFs and rejecting sources with measured
radii that were too small (\eg bad pixel, cosmic rays) or too large
(\eg galaxies, flat-field variations).  Stationary objects, defined as those
appearing at the same location in three or more images, were also removed
from the detection list.

\section{Object Identification}
\label{mops}

Moving objects were identified and linked by the Pan-STARRS telescope's
Moving Object Processing System (MOPS).  The heart of the MOPS detection
linking code is a variable kd-tree algorithm that allows the depth of the
kd-tree branches to be dynamically modified to increase search efficiency
\citep{kubica}.  This reduces the time requirements for linking large numbers
of detections in moderate-to-high noise data sets.  Orbits were computed
using the techniques of \citet{milani} as implemented for MOPS.  Our search
was restricted to only those objects moving with Main Belt-like rates of
motion.

We identified $828$ Main Belt objects with magnitudes brighter than
$g'\sim22.5$.  All detections for these objects were submitted to the Minor
Planet Center and 333 had no previous reported observations.  MPC
designations for all objects, as well as orbital elements, lengths of orbital
arc, number of observations, estimated albedos and derived diameters, are
listed in Table~\ref{tab.orb}.

\begin{figure}
\centering
\includegraphics[angle=-90,width=\textwidth]{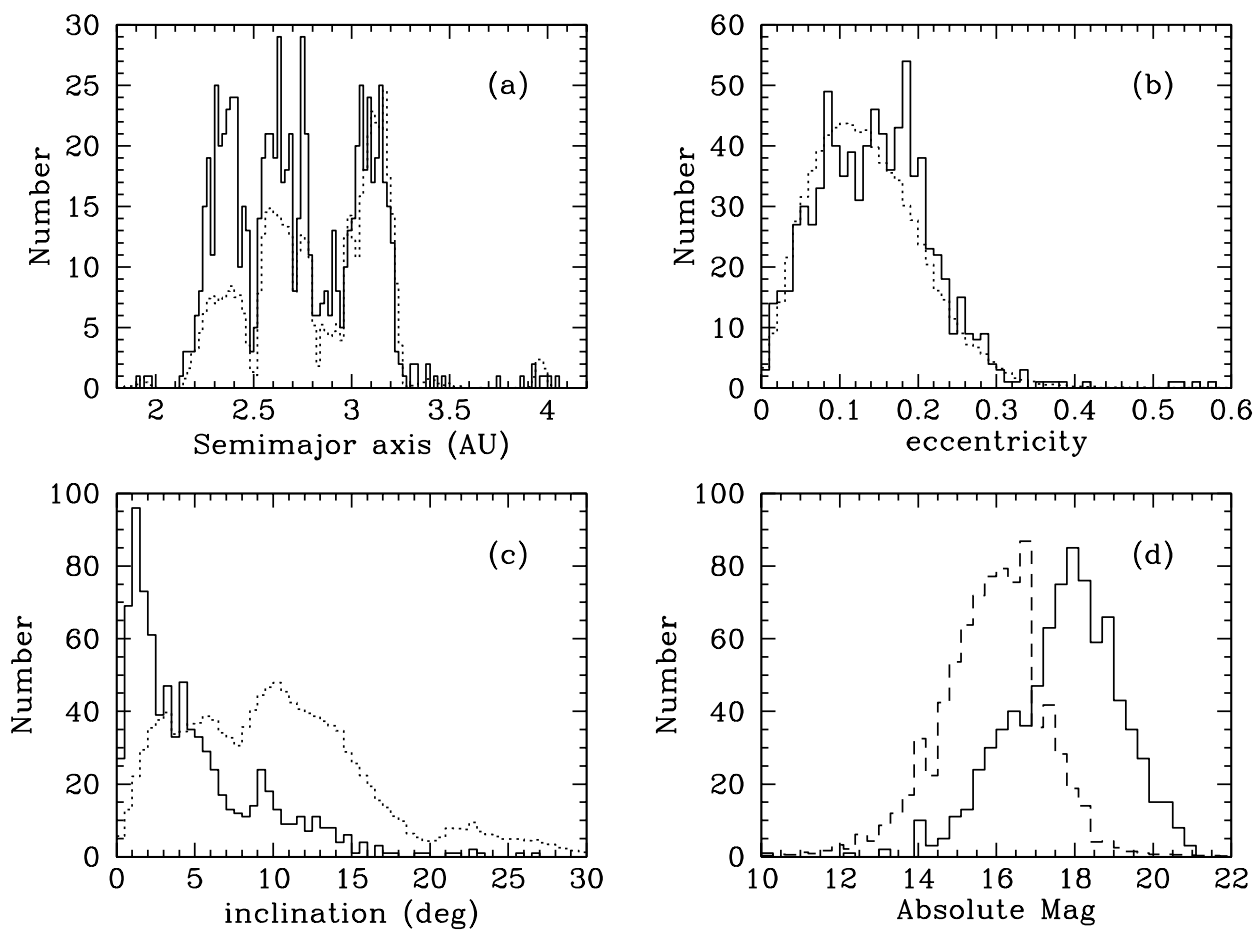}
\protect\caption{ The solid lines show the number distributions of
TALCS asteroids for a) semimajor axis ($0.02~$AU bins), b)
eccentricity ($0.01$ bins), c) inclination ($0.5^\circ$ bins), and d)
absolute magnitude ($0.3~$mag bins).  The dotted lines show the
arbitrarily normalized distributions of all known Main Belt asteroids
with $H_v<15$ for comparison while the dashed line is the arbitrarily
normalized distribution for all known Main Belt asteroids.  }
\label{fig.hists}
\end{figure}

\begin{figure}
\centering \includegraphics[scale=0.7]{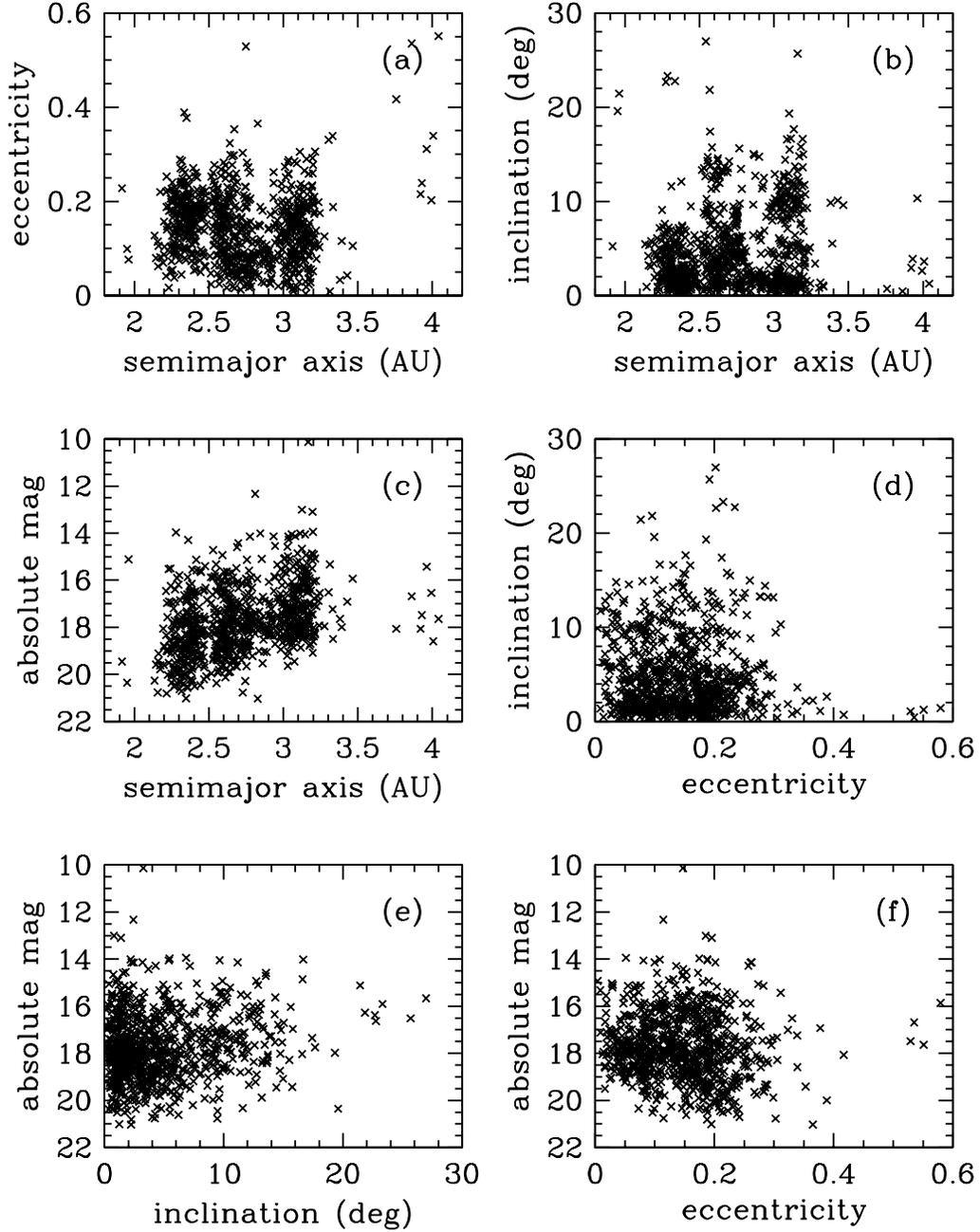}
\protect\caption{ Scatter plots for the TALCS objects of: a)
eccentricity vs semimajor axis, b) inclination vs semimajor axis, c)
absolute magnitude vs semimajor axis, d) inclination vs eccentricity,
e) absolute magnitude vs inclination, and f) absolute magnitude vs
eccentricity.  }
\label{fig.scatter}
\end{figure}

Figure~\ref{fig.hists} shows number distributions of the semimajor axes,
eccentricities, inclinations and absolute magnitudes of the TALCS asteroids
compared to the known distribution from the Minor Planet Center
(MPC)\footnote{\it http://www.cfa.harvard.edu/iau/lists/MPDistribution.html}.
For comparison to the currently known population of asteroids, the dotted
lines show the arbitrarily normalized distributions of semimajor axes,
eccentricities and inclinations for all asteroids with $H_v<15$ (\ie a
``complete'' distribution) while the dashed line shows the normalized
absolute magnitude distribution for all asteroids known.  The TALCS semimajor
axis distribution shows that the survey has nearly equal sampling from the
inner-, mid-, and outer-Main Belts, in contrast to the actual distribution of
objects with semimajor axis.  This is the result of inner Main Belt asteroids
being brighter and thus easier to identify in a magnitude-limited
survey. Figure~\ref{fig.hists} also shows that the TALCS objects'
eccentricity distribution is very similar to the MPC data, while the
inclination distribution shows a strong preference for low inclination
objects, as is expected for a survey restricted to the ecliptic over a short
time span.  The TALCS absolute magnitude distribution peaks about two
magnitudes fainter than the MPC data reflecting our focus on small asteroids.
The scatter plots in Fig~\ref{fig.scatter} show no serious gaps in the
orbital element coverage of TALCS, with the semimajor axis vs absolute
magnitude figure showing the effects of the apparent magnitude limit of our
survey with the lack of faint, distant objects.

\begin{figure}
\centering \includegraphics[scale=0.6]{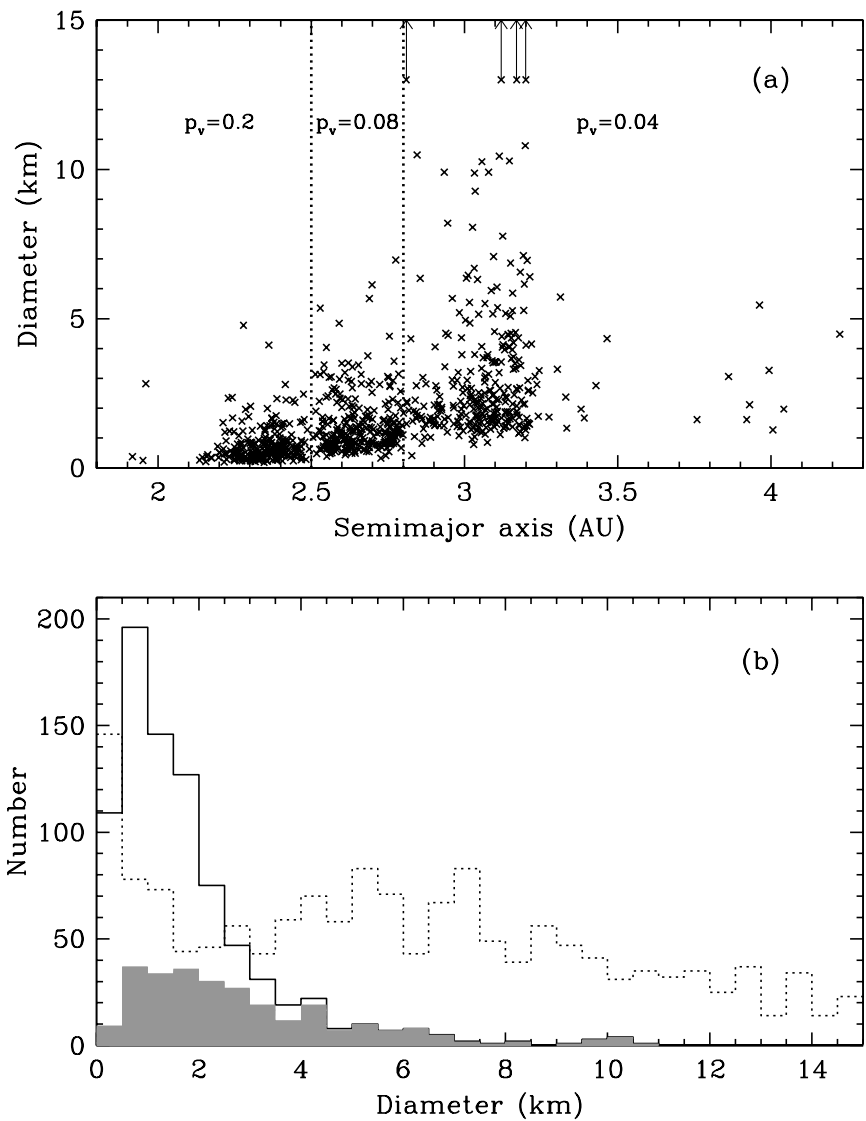}
\protect\caption{ a) The derived diameter vs. semimajor axis for the
TALCS population.  Dotted lines indicate the semi-major axis ranges
for assigning albedos ($p_v$).  Four asteroids in our survey ((755)
Quintilla, (3186) Manuilova, (4863) Yasutani, and (8906) Yano) have
diameters beyond the range of the figure and are indicated by x's with
arrows at the appropriate semimajor axis.  b) The solid line provides
the number distribution for all TALCS objects as a function of their
diameter ($0.5~$km bins) while the shaded region shows the
distribution only for those TALCS objects with measured light curves.
The four objects with $D>15~$km are not shown in this histogram.  The
dotted line shows the diameter distribution of all objects with
measured rotation periods and $D<15~$km as compiled by A.W. Harris,
et~al. (\it{http://www.minorplanetobserver.com/astlc/LightcurveParameters.htm}).
}
\label{fig.aDscat}
\end{figure}

Albedos for all objects were estimated based on their heliocentric distance
following the known decrease in albedo with distance from the Sun
\citep{tedesco05}.  The semimajor axis ($a$) ranges for assumed albedos
($p_v$) were: $a\le2.5~$AU, $p_v=0.20$; $2.5<a\le2.8~$AU, $p_v=0.08$;
$a>2.8~$AU, $p_v=0.04$, as shown in Fig~\ref{fig.aDscat}.  Diameters were
then calculated using:
\[D=\frac{1329}{\sqrt{p_v}}10^{-H/5}\]
\citep[see][and references therein]{harrisAIII}.  Figure~\ref{fig.aDscat}a
shows that TALCS is sensitive to objects as small as $1~$km diameter through
the entire Main Belt with much better sensitivity to smaller objects at
smaller semimajor axes as expected in a magnitude limited survey.  We compare
our diameter distribution for all TALCS asteroids and for only the light
curve fit objects to the diameter distribution of the smallest objects with
known periods\footnote{compiled by A.W. Harris, et al. in November 2008;
available online:
\it{http://www.minorplanetobserver.com/astlc/LightcurveParameters.htm}} in
Fig~\ref{fig.aDscat}b.  The known objects show a flatter diameter
distribution than the TALCS data indicating that TALCS is preferentially
sensitive to smaller Main Belt asteroids.  Asteroid (755) Quintilla was the
only TALCS object observed in the IRAS survey and has a derived diameter of
$36.0\pm2.1~$km \citep{tedesco02}.  This is significantly lower than our
estimate of $\sim62~$km, however it was found to have an unusually high
albedo of $p_v=0.1621$ for an object with $a>3~$AU.


\section{Light Curve Analysis}
\label{lcfit}

The TALCS observing program was designed to have sensitivity to light curve
periods ranging from as short as $1$ hour to over $50$ hours.  This was
accomplished by varying the transient time interval (TTI; the time between
repeated exposures on the same field) from $2$ to $17$ minutes as listed in
Table~\ref{tab.survey}.  The observations were spread out over nights with
different spacings to reduce the effect of aliasing and extend the period
sensitivity window.  Because of this survey design the number of points in
the light curve of different objects varied from $20$ to $140$ with a maximum
of four hours of consecutive observations on any single night.  This placed
our light curves in a difficult regime between the typical dense data sets
generated by single-object light curve surveys and the sparse light curve
data that will be generated by the next generation surveys like Pan-STARRS
\citep{mikko04,durech07}.

\begin{figure}
\centering
\includegraphics[angle=-90,width=\textwidth]{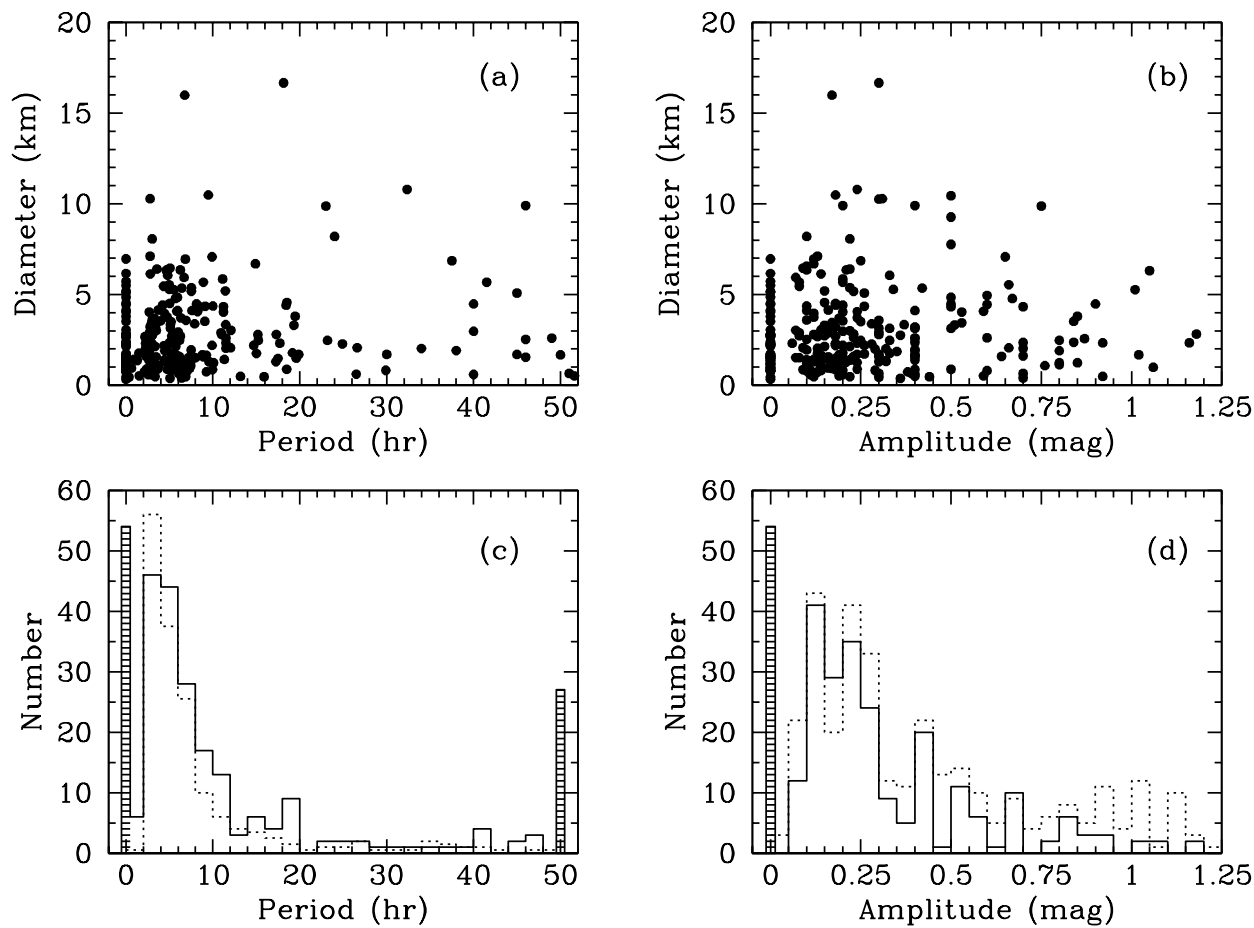}
\protect\caption{ a) Diameter vs. period for all light curve-fitted
objects.  b) same as (a) but vs. light curve amplitude (mag).  c)
Number distribution of fitted periods of TALCS objects ($2~$hr bins).
The shaded bar at $P=0~$hr represents all objects with no detectable
light curve while the shaded bar at $P=50~$hr includes all objects
with $P>50~$hr.  d) The number distribution of light curve amplitudes
($0.05~$mag bins).  The shaded bar at $P=0~$hr represents all objects
with no detectable light curve.  The dotted line in (c) and (d) shows
the distribution of previously known light curve periods and
amplitudes for asteroids in the same size range as the TALCS objects.
}
\label{fig.lcfits}
\end{figure}

Light curve periods were fit to each data set that was of sufficient
photometric quality.  The statistical and systematic errors of the photometry
established a limiting magnitude for the fits of $g'\sim 21~$mag
corresponding to S/N$\sim 15$.  We fit periods and amplitudes to $278$ TALCS
asteroids using the Fourier series method that was developed by
\citet{harris89} and has since become the standard for light curve fitting
\citep[\eg][]{pravec05}.  Figure~\ref{fig.lcfits} shows that no objects
larger than $D=2~$km were found to have periods less than $2~$hours.  (The
cluster of objects with $P=0~$hr are those that show no significant
photometric variation with rotation.)  Other than this there is no strong
relationship observed between diameter and either period or amplitude.  The
bulk of the TALCS objects have periods between two and ten hours and
amplitudes less than $0.4~$mag, though a number of objects were found with
periods up to and above $50~$hours.  Few objects were found with amplitudes
at or above $0.75~$mag which would indicate extreme shapes not typical for
relaxed bodies.  The error in the amplitude determination for many objects
was of order $0.1~$mag, causing a cluster of objects at \eg $0.4~$mag,
$0.5~$mag and $0.7~$mag values.  These peaks in the amplitude distribution
should be considered observational artifacts though the overall trend in the
distribution is real.  Comparisons to the previously known light curve data
are shown for both distributions and confirm that prior to debiasing TALCS
shows the same general trends as seen in earlier surveys.  The major
exceptions to this are objects with very short ($P<2~$hr) and very long
($P>30~$hr) rotation periods.  This result is discussed in \S~\ref{results}.

A full list of period and amplitude fits as well as $g'-r'$ color and
reliability parameter (U) is given in Table~\ref{tab.lcfits}.  The
reliability parameter follows the definition by \citet{harris83} with a
modification to the ``0'' value:
\begin{itemize}
\item [0]: no observed variation beyond photometric error
\item [1]: fragmentary or inconclusive coverage, possibly wrong
\item [2]: fairly conclusive result, may be incorrect at the $10-20\%$ level or a multiple (\eg $0.5$, $2$, etc) of the true period
\item [3]: secure result with essentially all of the rotation phase covered
\item [4]: multiple opposition coverage with pole estimation
\end{itemize}
Due to the nature of the TALCS data set and the absence of overlap with any
previous light curve survey, all results presented here are single-opposition
light curves and thus none will have a reliability parameter of U$=4$.

\begin{figure}
\centering
\includegraphics[angle=-90,width=\textwidth]{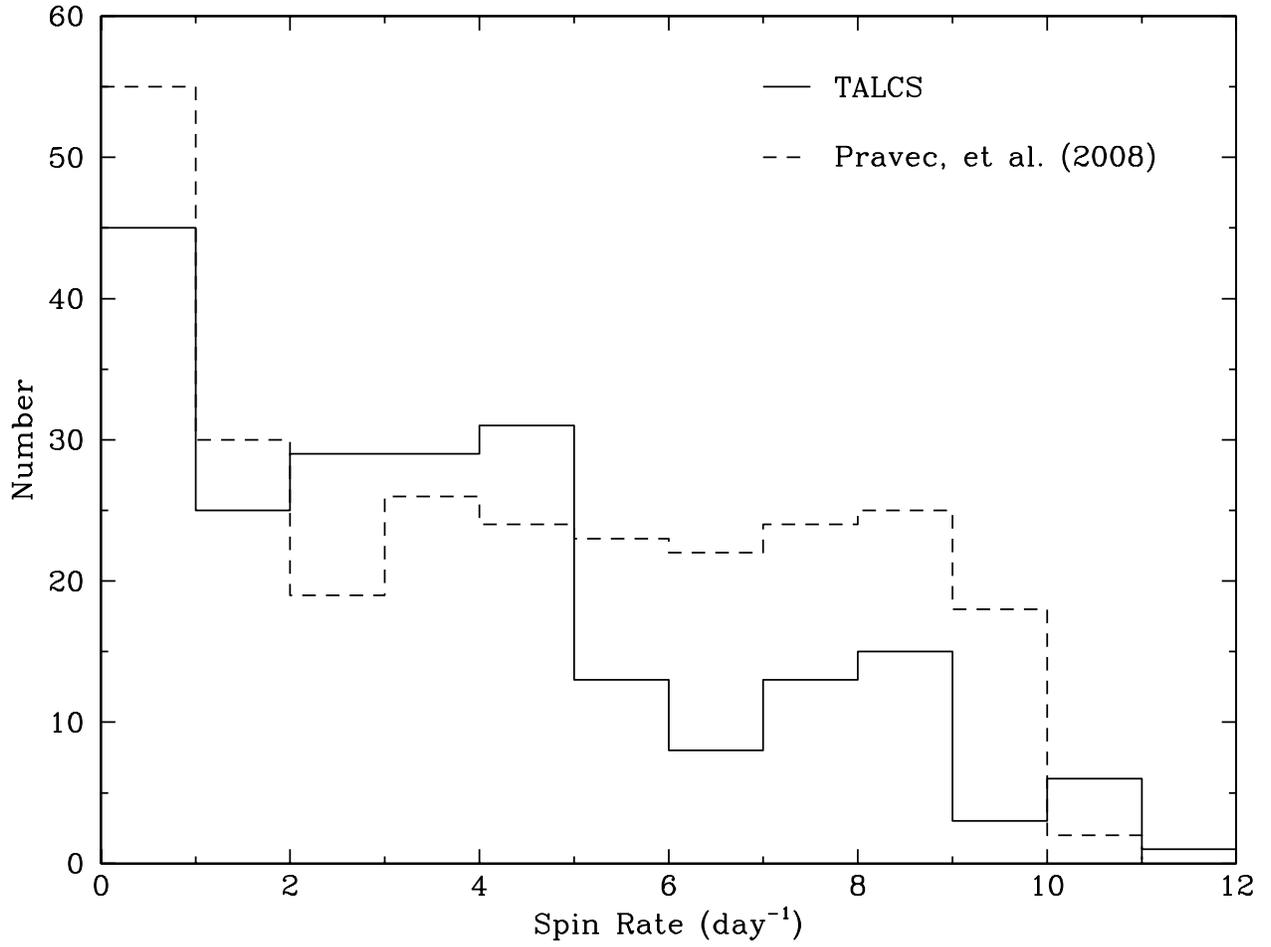}
\protect\caption{ Raw spin rate distribution for TALCS objects (solid)
compared to the data presented by \citet[][dashed]{pravec08}.  }
\label{fig.rate}
\end{figure}

The raw spin rate distribution for TALCS asteroids in Fig~\ref{fig.rate}
reveals a decrease in the number of objects with increasing spin rates in
contrast with recent results presented by \citet{pravec08} who found a nearly
flat spin rate distribution.  We discuss the implications of these
contrasting results in \S~\ref{results}.

\section{Debiasing}
\label{debiasing}

TALCS was designed to minimize the influence of biases found in targeted
surveys such as favoring brighter and closer asteroids but was still
susceptible to three serious and unavoidable biases: 1) a decreasing
sensitivity to low amplitude rotators for fainter objects in our magnitude
limited survey, 2) fits becoming unconstrained as the rotation period becomes
a significant fraction of the total survey time window, and 3) data-loss due
to objects falling on chip gaps or moving out of the survey area.

For any set of light curve observations that do not have complete continuous
coverage of an object's full rotation, underlying periodicities in the data
can be mistaken for the rotation period, especially for weakly varying
objects.  This aliasing of survey periods into apparent rotation periods is a
major source of error in light curve surveys.  Examples of aliased periods
include: an apparent $24~$hour period arising from observing a target the
same time every night; periods on the order of a few minutes as dictated by
the time between subsequent exposures; integer multiples of the actual period
due to incomplete coverage.  Although the TALCS survey cadence was designed
to avoid aliasing problems, they can arise if one or more nights of data were
lost due to objects falling off of chips or out of the field of view.
Additionally, objects with light curve amplitudes comparable to the
photometric noise are difficult or impossible to fit correctly.

To measure these biases we generated a population of synthetic asteroids and
used automated light curve fitting software to determine our fitting
efficiency over a range of light curve periods and amplitudes.  To validate
the rotation periods from the TALCS survey we obtained followup observations
of a subset of objects.

\subsection{Synthetic Light Curves}

We generated $100,000$ synthetic light curves including realistic photometric
noise over a range of periods, amplitudes, and magnitudes spanning the values
found in TALCS.  The rotation periods and amplitudes were generated with a
flat distribution over the ranges: $1~$hr $< P < 30~$hr, $0.05~$mag $< A <
1.2~$mag.\footnote{Although correlations in rotation states have been
observed for asteroids of the Koronis family \citep{slivan03} no such
relationship has been found for the Main Belt as a whole and we found no
evidence of any correlation in our raw TALCS data.}  To ensure that the
synthetic objects' apparent magnitude distribution matched the TALCS objects'
apparent magnitude distribution we randomly generated the synthetic objects'
magnitudes from a fit to the observed magnitude ($M$) distribution of the
form:
\[N=\frac{2.5^{M-18.6}}{1+e^{(M-20.6)/0.25}}\]
This function simultaneously accounts for the increase in the number of
objects with apparent magnitude as well as the falloff in detection
efficiency inherent to a magnitude-limited survey \citep{jedicke97}.

Assuming the synthetic objects are relaxed triaxial ellipsoids, the $b$ and
$c$ axes are equal, and the asteroids are in a principal-axis rotation state,
their light curves are described by:
\[m=2.5\log_{10}\sqrt{1+\left[\left(\frac{b}{a}\right)^2-1\right]\cos^2(2\pi\phi)\sin^2\theta}\]
and the amplitude of the light curve ($\Delta m$) is:
\begin{eqnarray}
\Delta m&=&2.5\log\left[\cos^2\theta+\left(\frac{b}{a}\right)^2\sin^2\theta\right]^{-1/2} \label{eqn.dmag}
\end{eqnarray}
where $m$ is the relative magnitude, $b/a$ the ellipsoidal axis ratio,
$\theta$ is the angle of the spin vector with respect to the line of sight
and $\phi$ the rotation phase \citep{pedro03,pedro07}.  The orientation of
the poles of the synthetic asteroids as well as the initial phases were
isotropically distributed across the full range of values.

Synthetic photometric measures were drawn from the light curves using the
TALCS observing cadence.  Data loss for single observations and whole nights
was simulated based on the actual data loss rate for the real objects.  The
most common reason for loss of a single observation was passage through a
star or diffraction spike while loss of a whole night occurred when objects
moved into chip gaps.

The Fourier series method of light curve fitting requires significant user
interaction and is not a feasible method for fitting the light curves
generated from the $100,000$ synthetic objects.  Instead we used an adapted,
simplified version of the light curve inversion method presented in
\citet{mikko07}.  This technique uses the full data set to constrain all
possible periods for a range of triaxial ellipsoid shapes that could be
responsible for the observed light curve to determine the best-fitting period
and shape solution.  Due to our survey covering only a single opposition the
shape/pole solution is unconstrained but the period solution is usually good.
\citet{durech07} used this same technique to efficiently determine light
curve periods from synthetic data of a simulated 10-year Pan-STARRS survey.
Amplitudes for our synthetic light curves were then determined using the
best-fit periods to fit sinusoids to the data.

It must be noted that the use of triaxial ellipsoids to generate the
synthetic data followed by restricting the solution to be a triaxial
ellipsoids will contaminate the final efficiency measurement as the real
TALCS asteroids are almost certainly not perfect triaxial ellipsoids.  Due to
the artificial restrictions this puts on the fit parameters it is likely that
the efficiency measurement if incorrect will be too large, especially for
low-amplitude objects however an underestimation of the efficiency is not
ruled out.  To quantify the systematic impact of our technique we would need
to generate and fit a series of different synthetic shapes based on a variety
of physical models.  Alternate methods for generating synthetic asteroid
shapes and light curves include ``genetic'' combination of currently known
shapes \citep{mikko07} and randomized Gaussian-sphere models
\citep[\eg][]{mui96,mui98,vok02}.  While both methods would remove some
uncertainty created when restricting the fit to the type of shape generated
practical considerations prohibit their use for the current investigation.
Both require finite element modeling to describe shapes and applications of
scattering theory at each time step to generate light curves for each of the
$100,000$ synthetic object, a task beyond the scope of this current work.
Future investigations using these models will be conducted however for the
present paper we will restrict ourselves to the simplified model.

We restricted the range for period fitting to $1-30$ hours to reduce the
number of wrong solutions and computation time.  The lower limit was set to
prevent false solutions that arise from the aliasing of the $\sim15~$min
observing cadence.  The upper bound was required to reduce the processing
time for the synthetic survey to a reasonable level while maintaining good
coverage of the real objects' rotation periods.  Most real objects with
$P>30~$hours had large error bars on the period and increasing the range to
\eg $1-40~$hours would only increase the number of real objects contained in
the range by five: less than $2\%$ of the total sample.

\begin{figure}
\centering \includegraphics[scale=0.7]{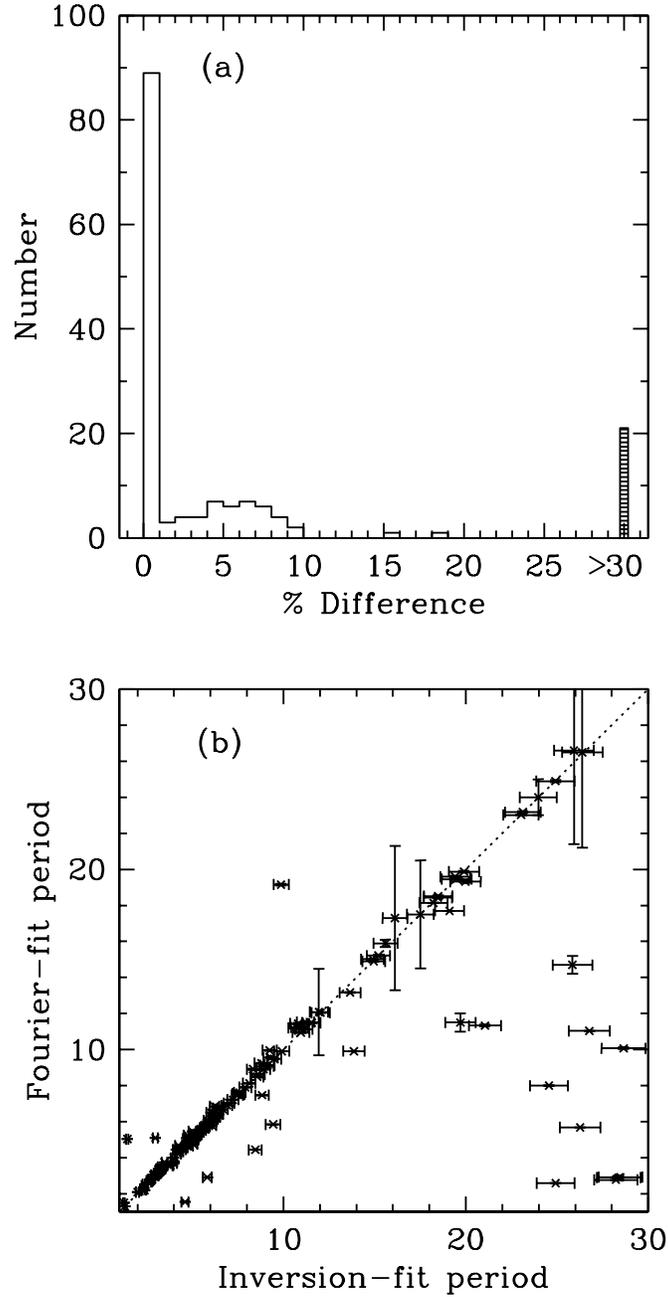}
\protect\caption{ a) Percent difference between the Fourier-fit and
Inversion-fit periods.  The shaded bar at $30\%$ represents all
objects with errors greater than this value.  The median error was
$\sim0.2\%$.  b) Comparison between periods found for the Fourier-fit
light curve fitting method and the Inversion (automated) fitting
technique over the period range to be debiased.  Errors on the
inversion periods were set to the $1-\sigma$ error level found in (a).  
}
\label{fig.LCscat}
\end{figure}

Starting with all TALCS objects that had Fourier-fit rotation periods between
$1-30~$hr we re-fit the observations using the light curve inversion method.
A comparison between the two methods is shown in Fig~\ref{fig.LCscat}.  The
top panel shows the percentage difference in the period between the two
methods where the shaded bar indicates all errors greater than $30\%$.  The
median difference was $0.2\%$ while a difference of $4.2\%$ encompasses
$\sim66\%$ ($\sim 1$-$\sigma$) of the sample.  We used the latter value as
the error on the inversion period as shown in the lower panel.  The automated
fitting of the real objects recovered the Fourier period to within a $4.2\%$
margin of error for over $85\%$ of the sample as shown in
Fig~\ref{fig.LCscat}a.  For the cases where there was a larger disagreement
between the two methods, we found that the Fourier periods were randomly
distributed while the inversion method periods showed a preference for
$\sim24~$hour.  This is likely the result of incomplete filtering of aliasing
due to the nightly cycle of observations.  In the remainder of this work we
use the Fourier-fitted periods for interpretation of the data and restrict
the use of inversion fitting to the determination of the observational
biases.

\begin{figure}
\centering \includegraphics[width=\textwidth]{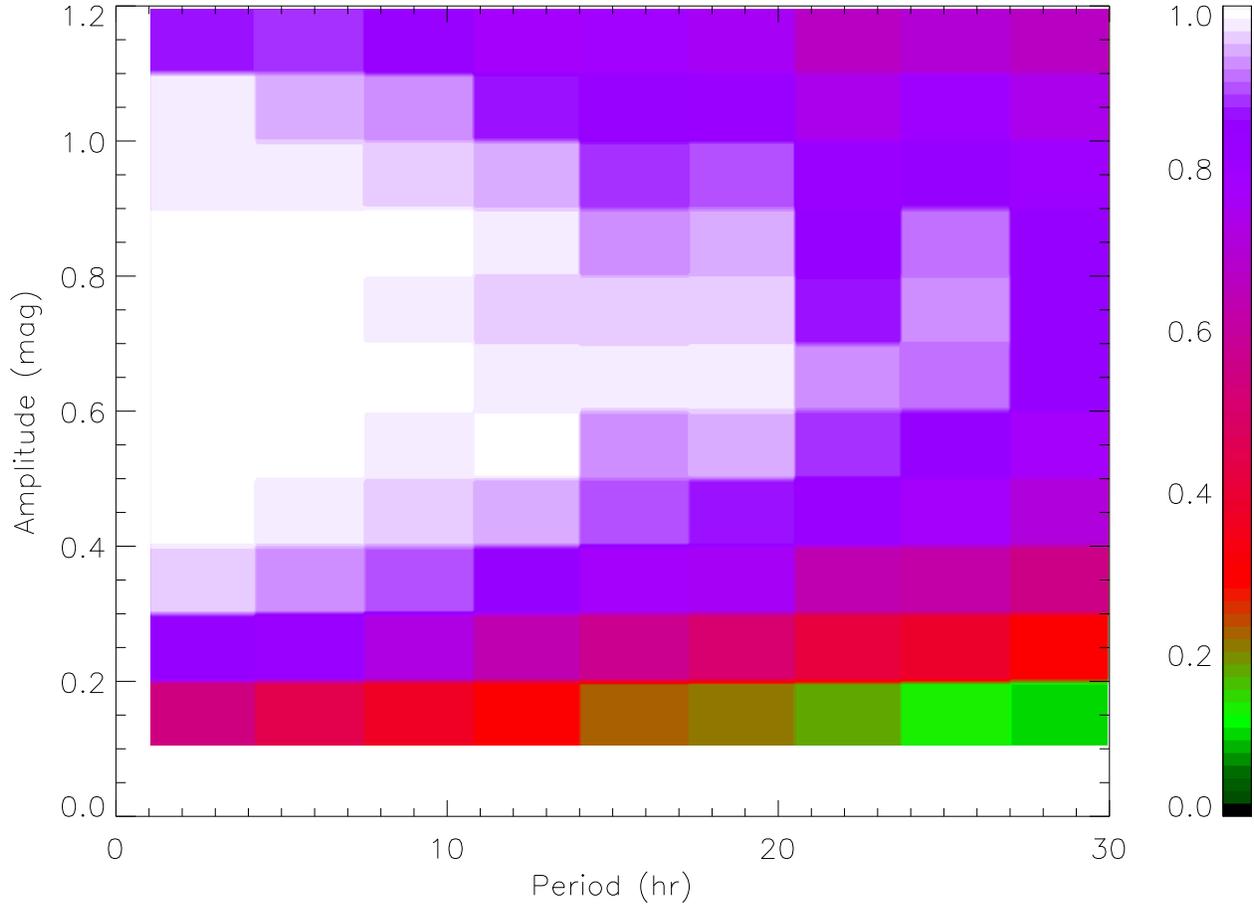}
\protect\caption{ Light curve fitting efficiency as determined from
our test using 100,000 synthetic objects.  The colorbar scale on the
right shows the fraction of correctly measured synthetic light curves.
Errors on the efficiency in each bin were less than $0.02$ in all
cases.  }
\label{fig.grideff}
\end{figure}

To debias the survey as a function of rotation period and amplitude we
divided the period-space into nine $3.25~$hour wide bins and the
amplitude space into eleven $0.05~$mag wide bins.  The number of
objects, $n_{ij}$, observed in bin ($i$,$j$) in the period-amplitude
space is given by:
\[ n_{ij} = \epsilon_{ij} N_{ij} + \sum_m \sum_n c_{ijmn} N_{mn} \]
where $N_{ij}$ is the number of objects generated in bin ($i$,$j$),
$\epsilon_{ij}$ is the efficiency of recovery in that bin, and $c_{ijmn}$ is
the crosstalk between bins: the fraction of objects generated in bin
($m$,$n$) that are recovered in bin ($i$,$j$).  In our simulation we found
that only $14$ of the $100,000$ objects were generated in one bin and
recovered in another, thus $c_{ijmn} \sim 0$ in all cases.  Nearly all the
objects that were not correctly identified failed the fitting process and
were placed in the $P=0$, $\Delta m = 0$ bin (which was not covered in our
debiasing).  Thus, we calculate the efficiency in each bin as:
\[ \epsilon_{ij} = n_{ij} / N_{ij} \]
Figure~\ref{fig.grideff} is a map of our light curve fitting efficiency in
period-amplitude space.  Objects with periods less than 20 hours and
amplitudes larger than $0.3~$mags are recovered with $90-100\%$ efficiency,
with the efficiency decreasing for longer periods and smaller light curve
amplitudes. Using these efficiencies we calculate the actual number of
objects in each bin from the number of detected objects and set upper limits
on the population for bins that do not contain real objects.  This results in
a grid of debiased populations for each period-amplitude bin.  We then
collapse these bins into distributions in period and amplitude space as
discussed in \S~\ref{results}.

While the determination of the efficiency from the synthetic models is
straight-forward, measuring the errors on that efficiency is considerably
more difficult.  Standard counting statistics (\eg Poisson statistics)
cannot be applied, as measurement of efficiency is not an inherently random
process.  Instead, we turn to the Jackknife method of error analysis used
frequently in cosmology where a plethora of synthetic models exist and must
be compared to a single sample, the Universe.  Jackknife measurements have
been used \eg~for determining covariance matrices in the large-scale
structure power spectrum \citep{pope08,henry08}, and a detailed mathematical
description of this methodology can be found in both \citet{gottleib} and
\citet{lupton93}.  In short, the jackknife method involves removing a subset
of the modeled population and recomputing the quantity of interest.  This is
done for all subsets and the variance in the computed quantity is then
calculated, modulo a normalization factor accounting for the population size
and subset size.  Typically, each unit in the model is considered a subset
and removed individually.  For our synthetic simulation we computed errors by
solving for the efficiency in each bin with each unit removed.  Without
exception we found that the errors in each bin due to the synthetic
population were significantly smaller than the errors due to the small number
statistics of the population of each bin.  For example, the largest jackknife
efficiency error on a single bin was $0.54\pm0.02$, while the bin with the
largest fractional error was $0.11\pm0.01$ or a $9\%$ error.  This is well
below the fractional errors due to the small numbers of objects in each bin
($\sigma\ge15\%$), meaning that our synthetic model does not dominate the
error on the final debiased value.  Both jackknife and small-number errors
are included in the final error for each debiased bin population.

\subsection{Followup Observations}
\label{followup}

To further validate our light curve determination we obtained followup
observations of a subsample of our TALCS targets.  Using the Tektronix
$2k\times2k$ CCD on the University of Hawaii's 2.2 meter telescope located on
Mauna Kea we reobserved $10$ asteroids over two runs in January and March of
2008.  Followup observations occurred when all targets were fainter than when
initially observed and this, coupled with using a smaller telescope, meant
that photometric errors were sometimes larger than in the TALCS data yielding
less precise periods.  The change in cadence between surveys, however, meant
that aliasing problems in TALCS could be identified.

For seven of the followup objects we obtained enough data to make light curve
determinations.  Starting from the period initially found from the TALCS data
we explored the surrounding period-space for the best-fitting period.  All 5
objects for which we obtained sufficient followup coverage to re-measure
their periods confirmed the original TALCS periods to within $2\sigma$ as
shown in Table~\ref{tab.newobs}.  The followup observations indicate that the
$U$ parameters for short period rotators from the initial CFHT data are
correct, while objects initially determined to have very long periods
($P>30~$hr) do indeed show rotation over long timescales.  Amplitudes from
the followup cannot be compared to those measured in the CFHT data as changes
in pole orientation will alter light curve amplitude and shape.

\section{Results}
\label{results}

\begin{figure}
\centering
\includegraphics[angle=-90,width=\textwidth]{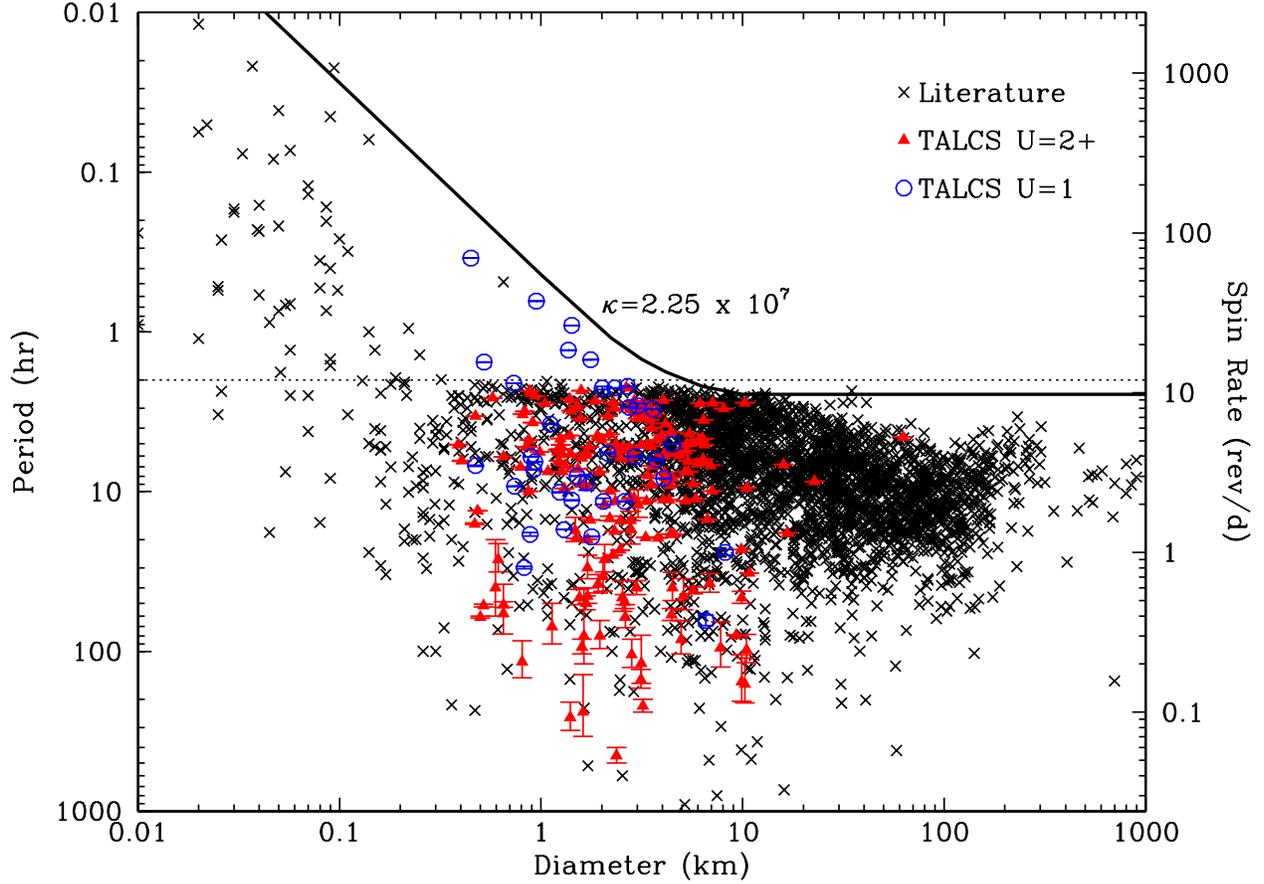}
\protect\caption{ Asteroid rotation periods and spin rates
vs. diameter.  The black x's are known values compiled by
A. W. Harris, et al.~current as of Nov 2008 ({\it
http://www.minorplanetobserver.com/astlc/LightcurveParameters.htm}).
Only light curves with a quality parameter (U) greater than 2 were
selected, resulting in a sample of 1442 objects.  The red filled
triangles are the TALCS data for all objects with fitted periods and
U$\ge2$ (287 Main Belt asteroids) while the open blue circles are the
U$=1$ TALCS objects (36 MBAs).  The dotted line shows a two-hour
period: the spin limit for a gravitationally bound body
\citep{pravec00}.  The thick solid line is the envelope from a
size-dependent strength with $\kappa=2.25\times10^7~$dynes cm$^{-3/2}$
reproduced from Fig.~5 of \citet{holsapple07}.  }
\label{fig.PvsD}
\end{figure}

Given the agreement between our two period fitting technique as shown in
Fig~\ref{fig.LCscat} and the agreement between the TALCS data and the
followup observations in Table~\ref{tab.newobs} we believe that the TALCS
data is suitable for studying the distribution of Main Belt asteroid rotation
rates and amplitudes.  Figure \ref{fig.PvsD} shows the Fourier-fit periods
and amplitudes for the TALCS objects compared to previously published
data.\footnote{compiled by A.W. Harris, et al. in November 2008; available
online:
\it{http://www.minorplanetobserver.com/astlc/LightcurveParameters.htm}} Those
data show a strong cutoff in period at $2.1~$hours for objects larger than
$\sim200~$m.  This rotation rate corresponds to the limit for a
gravitationally bound aggregate, or ``rubble pile'' \citep{pravec00}.  Above
this spin rate rubble pile asteroids should fission into binary systems with
both components having periods below this limit
\citep{scheeres07bin,walsh08}.  The single exception in the known data was
2001 OE$_{84}$ which was thought to be a unique circumstance of a very large
monolithic body \citep{pravecAIII}.

The TALCS results are distinct from the previously known data in two ways:
First, our survey detected six objects between $400~$m and $2~$km in diameter
with periods less than two hours --- faster than the critical spin rate for
objects in this size range.  Unfortunately, the accuracy of these spin rates
are questionable because the light curve amplitude for these objects is close
to the level of the photometric noise in the TALCS data sample.  At this
level the photometric noise may show random fluctuations at similar
timescales which could lead to a false period identification.  If these
objects are not rotating above the spin barrier, it would appear that the
gravitationally bound strength regime dominates Main Belt asteroids with $D
\ga 1~$km.

However, if any of these six fast rotators are confirmed by followup
observations they will support a size-dependent strength for Solar system
bodies as described by \citet{holsapple07}.  That work examined the effect of
a power-law distribution of cracks throughout a rocky object that induces a
falloff in tensile strength with increasing diameter.  Gravity stresses
dominate the strength of the largest asteroids ($D>10~$km) and a $\sim2~$hr
rotation limit exists even for bodies that are not rubble piles.  For smaller
objects the limiting critical spin rate ($\omega_{crit}$) increases as their
size decreases with $\omega_{crit}\propto~D^{-1.2}$.  \citet{holsapple07}'s
spin rate envelope encompasses both the largest objects rotating just beneath
the two hour `spin limit' as well as the smallest observed NEAs with rotation
periods of $P\sim1~$min.  All six of our objects with $P<2~$hr fall within
the envelope created by an assumed static strength coefficient of
$\kappa=2.25 \times 10^7~$dynes cm$^{-3/2}$ as shown in Fig~\ref{fig.PvsD}.
(The static strength, $k$, would then be $k=\kappa \bar{r}^{-1/2}$ where
$\bar{r}$ is the object's radius.)  From our debiased distribution we expect
that no more than $\sim4\%$ of Main Belt objects are in the superfast
rotation regime.

The second distinction between TALCS and previous surveys is the fraction of
objects found to have very long rotation periods.  We find in the TALCS data
$41$ asteroids with very long rotation periods ($P>30~$hr) out of $278$
objects with $D=1-10~$km.  The previously published data only contain $209$
objects with periods greater than $30~$hours out of $2669$ for all sizes
($87$ of these have diameters between $1-10~$km out of a total known
population of $871$ in that same size range).  This indicates that the biases
against long period objects in previous surveys are severe and that long
period objects are a significant fraction of the population of asteroids in
the Main Belt, especially at smaller sizes.  \citet{pravec08} find a similar
fraction of $P>30~$hr objects in their survey ($\sim18\%$).  Although the
Yarkovsky-O'Keefe-Radzievskii-Paddack effect (YORP) can be used to explain an
excess of slow rotators beyond that expected by collisional evolution
\citep[\eg][etc]{rubincam00,vok02,bottke06,rossi09} it is unclear whether
such a large excess of very slow rotators can be accounted for from this
effect alone.

\citet{vok07} show that as YORP slows the rotation of asteroids they fall
into a tumbling rotation state (non-principal axis rotation, or NPA).  TALCS
is not sensitive to NPA rotation, which can only be seen for data sets with
complete coverage of at least two rotation periods.  Additionally the
two-dimensional Fourier series fit used to characterize NPA rotation requires
an order of magnitude more data points than standard light curve
determination, up to $1000$ measurements for a high quality fit
\citep{pravec05} --- well beyond the TALCS data set.  The identification of a
large population of tumbling, slow-rotating asteroids would support a
YORP-driven model explaining the excess of slow-rotating objects.  A
competing possibility is that these objects have very high porosities which
would be very efficient at absorbing impacts, preventing energy transfer from
collisions to rotation.  This theory has been used to explain the extremely
large craters on (253) Mathilde as well as its slow rotation \citep[][and
references therein]{chapmanAIII} and could also explain the slow rotating
population in our data.  Additionally, as the TALCS data are not able to
distinguish binary asteroids from single objects the large excess of
slow-rotators could be close- or contact-binary objects with the observed
light curve variations indicating the orbital period \citep{harris02}.

\begin{figure}
\centering
\includegraphics[scale=0.7]{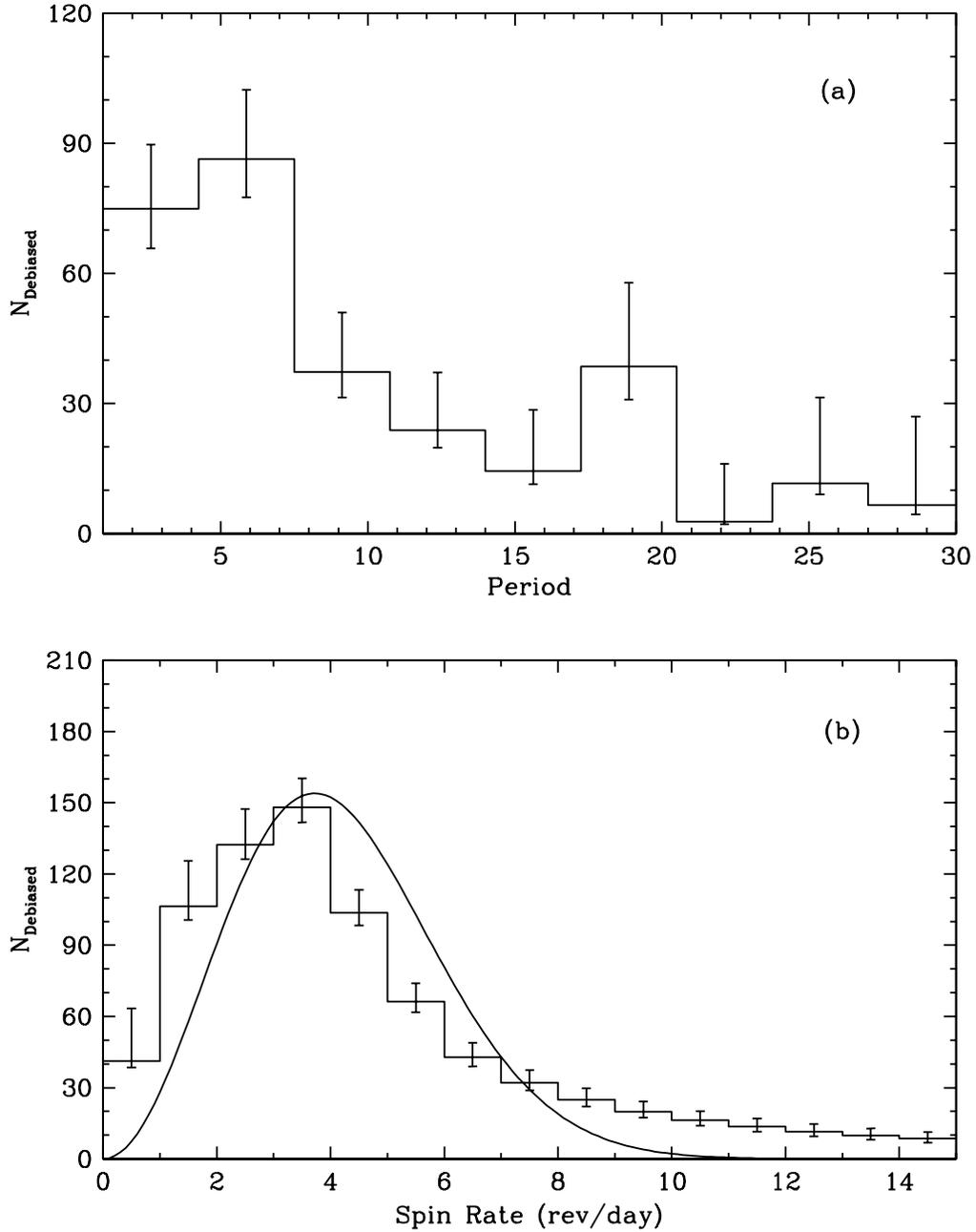}
\protect\caption{
(a) Debiased TALCS period distribution for the Main Belt ($3.25~$hr
bins). (b) Debiased TALCS spin rate distribution.  The solid curve is
the best-fit Maxwellian distribution with a mean rotation rate of
$4.19~$rev~day$^{-1}$.  }
\label{fig.ratePer}
\end{figure}

Using our calculated efficiency in period-amplitude space
(Fig~\ref{fig.grideff}) we debiased our distribution of light curves to find
the actual Main Belt period-amplitude distribution.  Collapsing that
distribution in amplitude yields the period distribution shown in
Fig~\ref{fig.ratePer}a while Fig~\ref{fig.ratePer}b shows the debiased spin
rate distribution as well as the best-fitting Maxwellian with a mean spin
rate of $4.19~$rev day$^{-1}$.  The Maxwellian was fit to a continuous
distribution of debiased spin rates but is shown in a binned differential
distribution as is common in the literature
\citep[\eg][etc.]{pravecAIII,pravec08}.  There are clear deviations from the
fit at both high and low rotation rates and these excesses cannot be
explained with collisional evolution alone but are likely due to the effects
of other processes such as YORP \citep{rubincam00} or binary breakup
\citep{harris02}.

There were $180$ out of $278$ ($\sim65\%$) of our period-fitted objects with
rotation periods in the debiased range of $1~$hr $ < P < 30~$hr.  Outside
this range, three objects ($\sim1\%$) have $P<1~$hr and $41$ objects
($\sim15\%$) had $P>30~$hr.  While the long period objects have uncertain
periods, all show clear light curve variation that is significantly longer
than $30~$hr and can be treated with confidence as a long-period group.
Finally, $54$ of our targets ($\sim19\%$) showed no variation in their light
curves above the photometric noise.  A lack of observable light curve will
result from one of three scenarios: 1) the asteroid has a shape that is
nearly a perfect sphere with no variation in albedo across the surface; 2)
the object is rotating with a period much longer than the survey window, in
the case of TALCS a multiple-month long period; or 3) the asteroid's rotation
pole is aligned very closely with the line of sight.  The YORP effect, used
to explain the slow-rotator population, breaks down when the rotation period
is a significant fraction of the orbital period \citep{vok02}, making it
difficult to create objects with multi-month periods.  It is also impossible
to explain this large population of flat light curves with pole orientation
alone without invoking an arbitrary and unphysical distribution of asteroid
rotation axes which should be isotropic due to collisional processing
\citep{salo87}.  Asteroid poles have been shown to be correlated for some
families \citep{slivan03} but this result cannot be applied to the Main Belt
as a whole.  Thus, a combination of shape and pole orientation are required
to explain this population of objects with flat light curves.

In their recent work with a similar size data set, \citet{pravec08} found a
flat spin rate distribution between $1$ and $9~$rev day$^{-1}$ for asteroids
with $3~$km $ \le D \le 15~$km and an excess of objects with spin rates less
than $1~$rev day$^{-1}$.  They cite the YORP effect as the cause of both the
excess of slow-rotating objects and the flattening of the spin rate
distribution.  The characteristic time to double or halve the rotation rate
of a $1~$km object is $\sim12-14~$Myr \citep{capek04} with approximately
equal numbers of objects accelerating and decelerating, though both the
amplitude and sign of YORP depend strongly on the shape of the asteroid
\citep{scheeres07yorp}.  The short timescales imply that the rotation rate
evolution of small Main Belt asteroids is dominated by YORP.  Its effect is
predicted to be independent of rotation rate until the object slows to
periods of hundreds of hours.  At very long rotation periods the current
models break down \citep{vok02,scheeres07yorp} and the object either remains
at a slow rotation rate or enters a tumbling state and evolves as a
non-principal axis rotator \citep{pravec05}.  In this way, YORP can be used
to explain some of the excess of slow rotators seen in both our data and that
of \citet{pravec08}.

\citet{pravec08} explain their flat distribution rate using the same YORP
models, arguing that the independence of YORP from the current rotation rate
leads to the erasure of any initial distribution function and results in a
flat rotation rate distribution.  They arrive at this conclusion using
simulations of YORP evolution.  In order to deal with the slow and fast
boundary conditions imposed by tumbling and breakup, respectively, they
reassign values of rotation rate changes when a boundary is reached.  Objects
slowed to $0~$rev day$^{-1}$ were reassigned a new value for change in spin
rate, while objects reaching the upper limit were wrapped to the same spin
rate in the opposite direction and allowed to slow from there.  If instead we
consider that objects with increasing rotation rates would eventually reach
the disruption limit, at which point they might disrupt into a binary or shed
mass \citep{scheeres07bin,walsh08}, the resultant shape change would lead to
changes in the amplitude and/or sign of the YORP effect on the body,
restarting its YORP evolution.  After many YORP-timescales have passed we
would find a population with a range of objects at different stages of
YORP-braking that started at various times.  This would lead to an increase
in the number of objects with decreasing rotation rate as observed in the
TALCS data.

We believe the difference in the measured spin rate distribution between
\citet{pravec08} and this work is due primarily to the differences in survey
methods.  TALCS was an untargeted survey while the Photometric Survey for
Asynchronous Binary Asteroids (BinAstPhotSurvey) from \citet{pravec08}
targeted individual objects.  A Kolmogorov-Smirnov test was used to compare
the two raw data sets and yielded a probability of $P=0.005$ that they were
drawn from the same population.  An important point is that TALCS samples
through the entire Main Belt while the BinAstPhotSurvey focused on the
inner-Main Belt where the YORP effect is more pronounced due to the relative
proximity to the Sun.

\begin{figure}
\centering \includegraphics[angle=-90,scale=0.7]{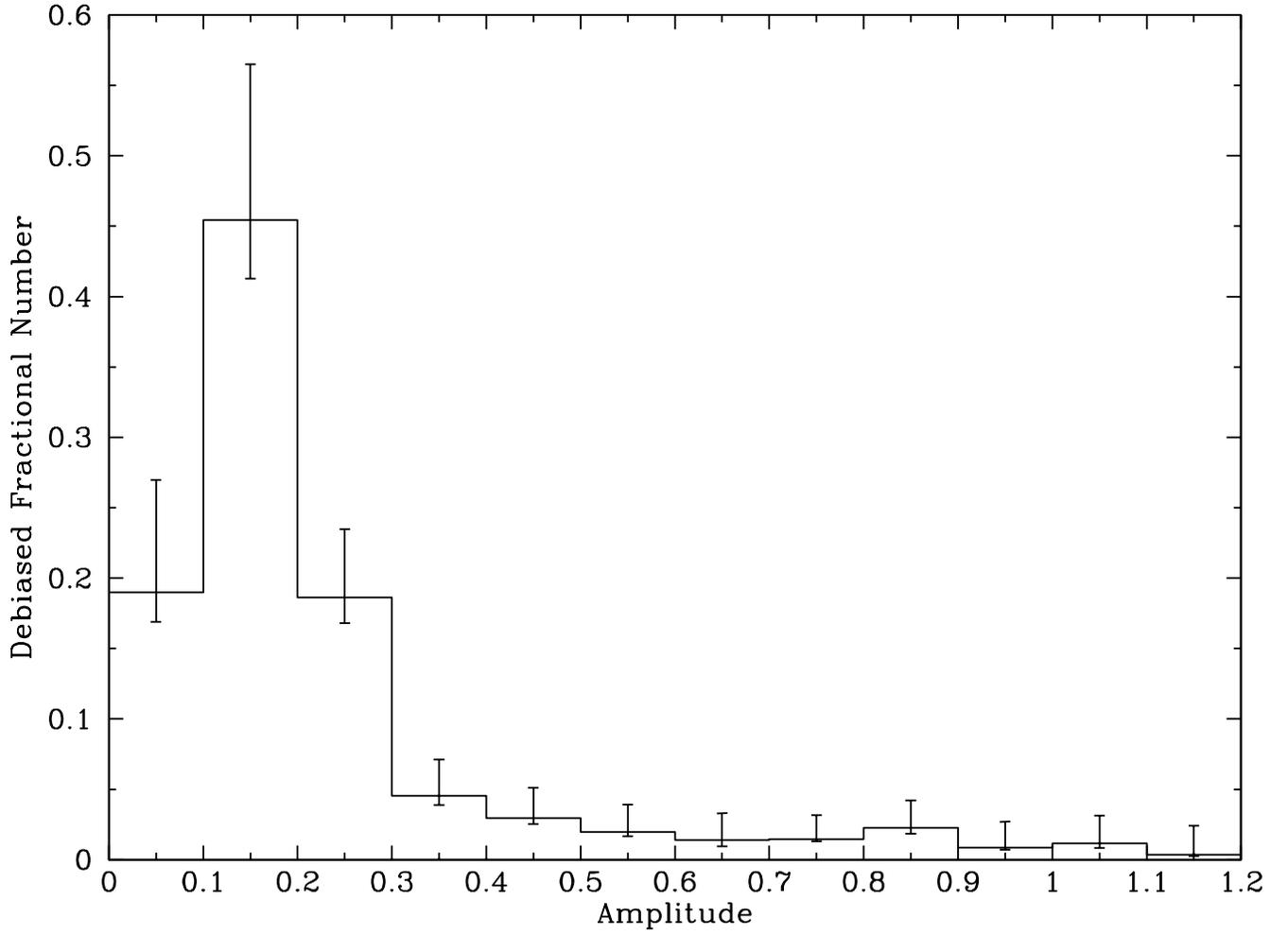}
\protect\caption{ Debiased fractional differential distribution of
light curve amplitudes for main belt asteroids.}
\label{fig.Dampdist}
\end{figure}

The binned differential distribution of debiased light curve amplitudes is
shown in Fig~\ref{fig.Dampdist}.  Most asteroid light curves have low
amplitude but there exists a long tail in the distribution such that a few
percent of asteroids have light curve amplitudes of $\ga 1$~mag, suggesting
that a similar fraction of asteroids are very elongated.

\begin{figure}
\centering \includegraphics[scale=0.6]{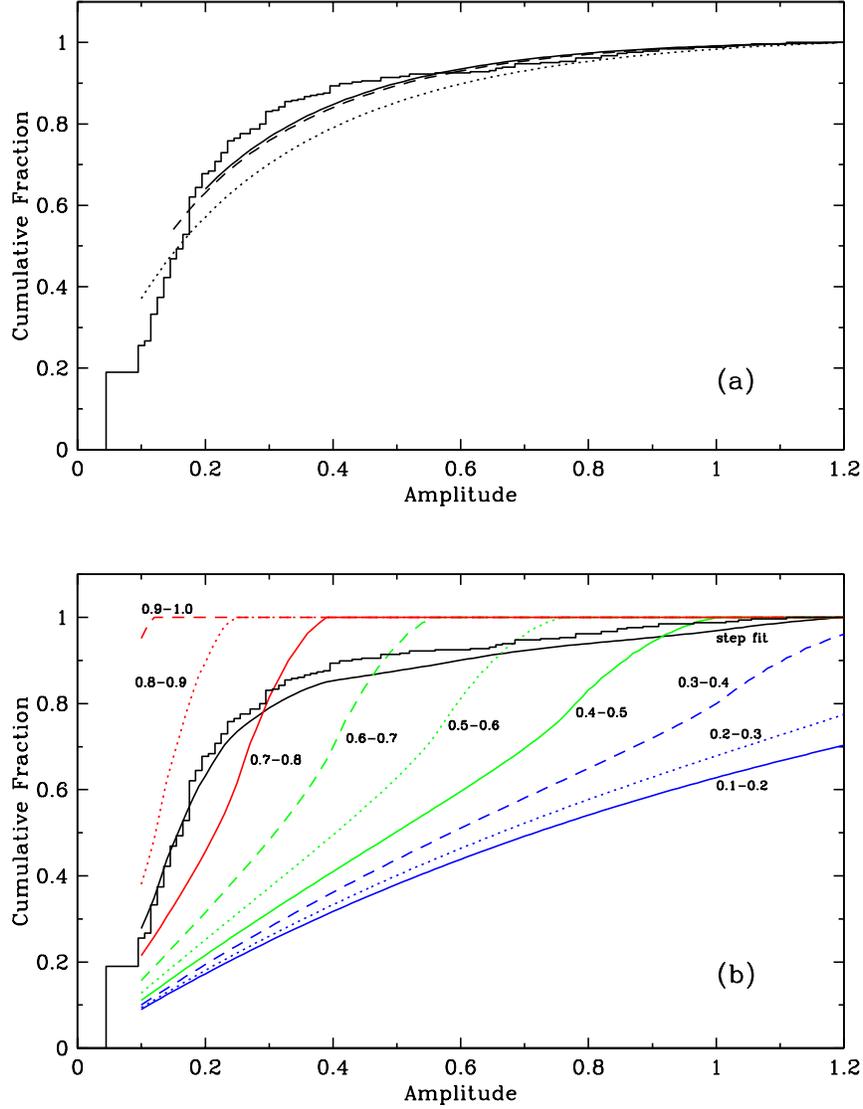}
\protect\caption{Debiased cumulative light curve amplitude distribution
for TALCS objects.  The amplitude$<0.1$ bin is an estimate of the
total number of objects with amplitude below this level based on the
number of objects in the survey with no apparent magnitude variation.
a) The smooth curves provide the best fits to the amplitude
distribution under the assumption that the $b/a$ axis ratios for the
underlying asteroid population is represented by a 2$^{nd}$ order
polynomial.  The dotted/dashed/solid curves correspond to cutoff
amplitudes (described in the text) of $0.1~mag$, $0.15~mag$ and
$0.2~mag$ respectively.  b) The solid black line labeled `step fit'
is the best fit to the amplitude distribution when we represent the $0
< b/a \le 1$ axis ratio distribution as a set of unconstrained 0.1
`steps' as explained in detail in the text.  The other curves
represent the shape of the cumulative amplitude distribution assuming
that all the objects fall into a single 0.1 wide $b/a$ ratio bin. }
\label{fig.Campdist}
\end{figure}

The unbinned cumulative distribution of the debiased light curve amplitudes
shown in Fig~\ref{fig.Campdist} was created by giving each real object a
weight based on the fitting efficiency of the bin in which it was located.
We assumed that the fraction of debiased objects with ``zero'' amplitude
(\ie $<0.1$~mag) was the same as the observed fraction of TALCS targets with
no amplitude variation ($19$\%).  This assumption has only a small affect on
the following analysis.

Assuming an isotropic distribution of rotation poles it is possible to
convert an asteroid shape distribution ($f(b/a)$) into a light curve
amplitude distribution or vice versa.  Following Eq~\ref{eqn.dmag} and
assuming random pole orientations we generated theoretical cumulative
amplitude distributions from different polynomial functional forms for the
shape distribution (testing orders 2, 3 and 4).  We required that $f(b/a)=0$
when $b/a=0$ and allowed the other polynomial coefficients to vary to obtain
the best fit to the debiased cumulative light curve amplitude distribution.
Our fitting metric was the `minimum greatest distance' between the generated
and debiased amplitude distributions, similar to a K-S test.  The second,
third and fourth order polynomials in $b/a$ yielded nearly identical fits so
we discuss only the second order result as shown in Fig~\ref{fig.Campdist}a.

The parameter with the strongest effect on the resultant shape distribution
was the smallest ``trusted'' amplitude.  As amplitudes decrease to within a
few sigma of the photometric noise even the most robust automated or manual
method of period fitting will begin to fit variations in the noise,
especially in data sets with non-continuous coverage like TALCS.  This
results in a large number of low-amplitude fits for fainter targets, similar
to the data from the \citet{dermawan} study.  To study and mitigate this
effect we reanalyzed the TALCS data cropping all amplitudes less than
$0.2~$mag (a level confirmed by our successful followup observations, see
\S~\ref{followup}).  The best fitting polynomial for the shape distribution
in each case is
\[\begin{array}{rcl}
\Delta m > 0.10 & : & f(\frac{b}{a}) = \left(\frac{b}{a}\right)^2 + 1.76 \frac{b}{a}\\
\Delta m > 0.15 & : & f(\frac{b}{a}) = \left(\frac{b}{a}\right)^2 + 0.07 \frac{b}{a}\\
\Delta m > 0.20 & : & f(\frac{b}{a}) = \left(\frac{b}{a}\right)^2 
\end{array}
\]
as shown in Fig~\ref{fig.Campdist}a. (We have suppressed a normalizing
constant that would guarantee that $\int_0^1 f(\frac{b}{a}) = 1$).  The best
fit is for $\Delta m > 0.2$~mag, where the greatest distance of
$dist=0.06$ can be compared to $dist=0.07$ for $\Delta m > 0.15$ and
$dist=0.13$ for $\Delta m > 0.10$.  The fact that the fit metric improves as
we increase the cutoff amplitude could be indicative of an overestimate in
the efficiency for fitting the lowest amplitude objects.\footnote{As a
further test, we performed the same analysis after removing all U=1 objects
--- those for which the periods and amplitudes are most uncertain.  There was
no significant change in the fit parameters and the result distribution in
the $b/a$ axis ratios were essentially identical.}  This could also indicate
that a simple polynomial function is not a good representation of $f(b/a)$.

To test this possibility we performed an `unparameterized' fit of the
cumulative light curve amplitudes to a normalized `step' distribution where
$f(b/a) = f_i$, $0.1(i) < b/a \le 0.1(1+i)$, for $i=0,9$.  We assumed that
the $b/a$ values were distributed evenly within each 0.1 wide bin and
integrated over all pole orientations.  The set of labeled smooth curves in
Fig~\ref{fig.Campdist}b shows the cumulative fractional distribution in light
curve amplitudes resulting from this test.  As expected, as $b/a$ increases
the power moves to higher amplitudes in the cumulative distribution.  We then
determined the combination of contributions from each single step
distribution that gave the best match to the observed distribution.  That
fit, the solid line labeled `step fit' in Fig~\ref{fig.Campdist}b, is better
($dist=0.05$) than any of the polynomial fits {\it and} included all light
curves with amplitude$>0.1~$mag.

\begin{figure}
\centering \includegraphics[angle=-90,scale=0.7]{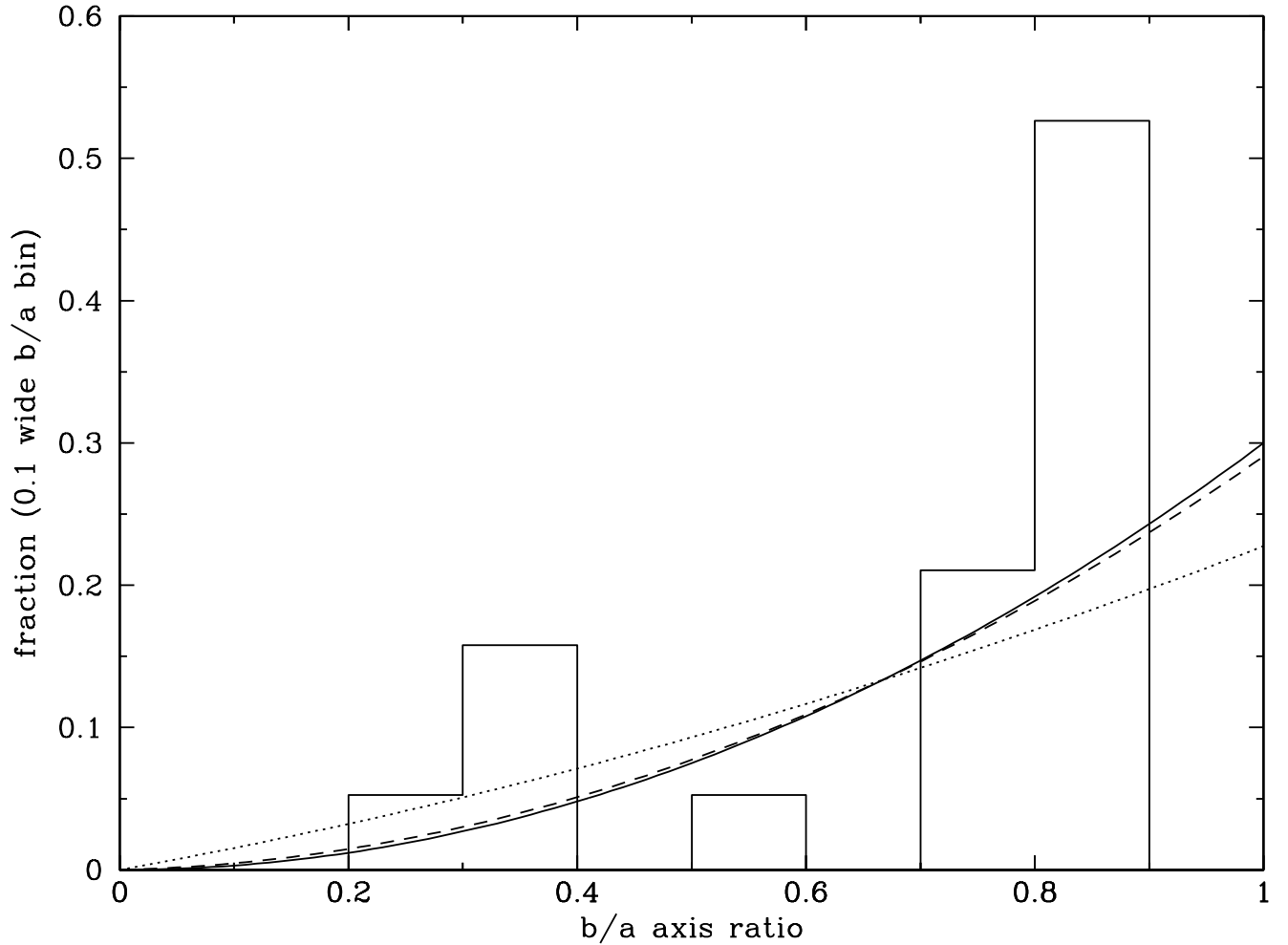}
\protect\caption{The fractional distribution of main belt asteroid
$b/a$ axis ratios from fits to the cumulative amplitude distribution
of Fig. \ref{fig.Campdist}.  The 3 smooth curves are 2$^{nd}$ order
polynomials with the same meaning as in Fig.\ref{fig.Campdist}: cutoff
amplitudes (described in the text) of $0.1~mag$, $0.15~mag$ and
$0.2~mag$ respectively.  The histogram is the result of a fit to the
cumulative amplitude distribution assuming that the $b/a$ axis ratios
of the asteroids can be represented `step wise' in 0.1 wide bins as
described in the text. }
\label{fig.badist}
\end{figure}

Figure~\ref{fig.badist} shows the $b/a$ probability distribution for the
polynomial and step fits.  Both types of fit are in general agreement in that
they suggest most main belt objects are closer to being spherical ($b/a \la
1$) than not.  Furthermore, the three polynomial fits with different minimum
amplitude cutoffs are roughly the same shape, two of which are essentially
indistinguishable.  The most interesting feature is the shape of the step fit
--- the fact that the quadratic polynomial is unable to fit the cumulative
light curve amplitude distribution and the relatively large fraction of
objects in the $0.3<\frac{b}{a} \le 0.4$ range of the step fit suggests a
bi-modality in the shape of these asteroids.

Since only small $b/a$ objects can produce large amplitude light curves, and
since we expect large objects to be more spherical due to gravitational
forces, we examined the diameter distribution for large and small amplitude
objects.  However, a K-S test between the diameter distributions of large
($>0.8$~mag) and small ($\le 0.8$~mag) light curve amplitude objects suggests
that they are drawn from the same intrinsic distribution --- at least within
the TALCS sample there is no reason to suggest that small objects are more
elongated than large objects in the size range sampled here.

\section{Conclusions}
\label{conc}

We present results from the Thousand Asteroid Light Curve Survey (TALCS), a
program designed to survey $12~$deg$^2$ of sky on the ecliptic and find
approximately $1000$ Main Belt objects in an untargeted manner using the wide
field MegaCam imager on CFHT and measure their light curves.  Using the power
of new software tools such as MegaPipe from the CADC and the Moving Object
Processing System from the Pan-STARRS telescope we have determined orbits and
photometry for all $828$ moving objects identified in the survey to a
$4~\sigma$ detection limit corresponding to a magnitude limit of
$g'\sim22.5$.  Of these, $278$ asteroids had photometry of sufficient quality
to fit multi-order sinusoids to the light curve and derive their rotation
period and amplitude.  Through a combination of survey design, cadence, and
synthetic light curve simulation we have debiased our sensitivity to both
period and amplitude in order to derive the actual light curve period and
amplitude distributions for the Main Belt.

We find that our debiased distribution of asteroids with rotation periods
between\\ $1~$hr $\le~P~\le~30~$hr can be roughly fit by a Maxwellian
distribution of rotation rates as expected for a collisionally evolved system
\citep{salo87} and as found for the largest ($D>40~$km) Main Belt objects
\citep{pravecAIII}.  However, there are strong deviations from a Maxwellian
at high and low rotation rates as reported by \citet{pravecAIII} for a mostly
Near Earth Object sample in the $0.15~$km $< D < 10~$km size range.  The
over-density of very slow rotators in the TALCS data exceeds that of the NEO
population.

The excess of slow-rotators becomes more pronounced including objects with
periods $P>30~$hr that fell outside our debiasing range.  We found that
nearly $15\%$ of our survey sample had periods greater than $30~$hours - a
much larger fraction than found in previous surveys
\citep[\eg][]{pravecAIII,pravec08}.  The discrepancy between TALCS and
previous work is probably due to the untargeted design of our survey with
wide observation spacing to allow sufficient coverage of long and short
period objects.

We find $6$ objects in our survey with diameters $D\ge400~$m that are
candidates for having rotation periods shorter than 2 hours. However, we note
that the periods for these objects are not well determined because of their
low amplitude light curves. At small amplitudes and short periods it becomes
difficult to disentangle noise fluctuations from the signal.  Previous works
\citep[][and references therein]{pravecAIII} have shown that an empirical
limit to rotation period exists at $\sim2~$hours that can be explained as the
rotation rate at which a gravitationally bound rubble-pile aggregate will
break up or begin shedding mass.  Objects with faster rotation rates must
have a non-zero internal strength holding the object together in addition to
gravity.  Prior to TALCS, only a single object in this regime was confirmed
\citep{pravecAIII} and was believed to be an unusual and unique object.  Our
six asteroids may represent a small population of Main Belt bodies larger
than $150~$m (the previously observed size limit for objects with
$P<2~$hours) that have some internal strength while our debiasing results
indicate this population to represent no more than $4\%$ of the Main Belt in
the $1-10~$km size range.  If these objects are confirmed during followup
observations they will lend support to the size-dependent strength model for
rocky bodies \citep[\eg][]{holsapple07}.

Finally, our fits to the debiased amplitude distribution ($\Delta m >
0.2$~mags) for objects with periods in the range $1~$hr $\le P \le
30~$hr indicates that the number distribution of asteroid shapes is
proportional to $(b/a)^2$.  Allowing smaller amplitudes produces worse
fits and the power of the axis ratio shape distribution moves away
from $b/a=1$.  A stepwise distribution in $b/a$ provides a superior
fit for $\Delta m > 0.1$~mag and suggests a large contribution
($\sim$75\%) from asteroids with round shapes ($b/a\sim0.8$) while the
remaining objects form a distinct group of elongated objects with
$b/a\sim0.3$.

\section*{Acknowledgments} 
The authors are most grateful for the period determinations of the light
curves provided by Petr Pravec.  We would like to thank the CFHT staff for
the hard work they put into scheduling and acquiring the complicated
observing cadence required by the TALCS program; in particular, Pierre
Martin, Mary Beth Laychak, Peter Forshay, and Adam Draginda.  We also would
like to thank Mikko Kaasalainen for his help adapting the automated period
determination software and for his careful review of the paper.  Many helpful
comments were also provided by Dan Scheeres
, Alan Harris (SSI), 
 and an anonymous reviewer.  J.M. was funded for this work under NASA PAST
grant NNG06GI46G.  J.\v{D}. was funded from GACR 205/07/P070 of the Czech
grant agency and Research Program MSM0021620860 of the Czech Ministry of
Education.  This research used the facilities of the Canadian Astronomy Data
Centre operated by the National Research Council of Canada with the support
of the Canadian Space Agency.  The authors wish to recognize and acknowledge
the very significant cultural role and reverence that the summit of Mauna Kea
has always had within the indigenous Hawaiian community.  We are most
fortunate to have the opportunity to conduct observations from this sacred
mountain.

\newpage

\begin{deluxetable}{cccccccc}
\tablenum{1}
\tabletypesize{\scriptsize}
\tablecaption{TALCS observing log}
\tablewidth{0pt}
\tablehead{
\colhead{UT Obs Date}   &
\colhead{Central RA}  &
\colhead{Central Dec}  &
\colhead{Pointings}  &
\colhead{Images per}   &
\colhead{Exposure}   &
\colhead{TTI$^a$}   &
\colhead{Filter} \\
\colhead{}   &
\colhead{(hh:mm:ss)}  &
\colhead{(dd:mm:ss)}  &
\colhead{}  &
\colhead{Pointing}   &
\colhead{Time (sec)}   &
\colhead{(min)}   &
\colhead{} 
}
\startdata
09-14-2006 & 01:12:22.58 & +07:40:08.9 & 12 & 17 & 30 & 15 & $g'$ \\
09-17-2006 & 01:10:17.19 & +07:27:26.2 & 12 & 10 & 40 & 17 & $r'$ \\
09-20-2006 & 01:07:58.57 & +07:13:19.4 & 2/6 & 35/30 & 20/30 & 2/8 & $g'$ \\
09-21-2006 & 01:07:09.68 & +07:08:19.9 & 12 & 18 & 30 & 15 & $g'$ \\
09-22-2006 & 01:06:19.55 & +07:03:12.3 & 12 & 17 & 30 & 15 & $g'$ \\
09-28-2006 & 01:00:56.75 & +06:30:01.4 & 12 & 6 & 30 & 15 & $g'$ \\
\hline
\enddata
\vskip 0.05in
\scriptsize{$^a$Transient Time Interval}
\label{tab.survey}
\end{deluxetable}

\begin{deluxetable}{crcccccccccccc}
\tablenum{2}
\tabletypesize{\scriptsize}
\rotate
\tablecaption{Asteroids identified in TALCS}
\tablewidth{0pt}
\tablehead{
\colhead{TALCS ID$^a$}   &
\colhead{MPC desig.}  &
\colhead{a}  &
\colhead{e}  &
\colhead{i}   &
\colhead{$\Omega$}   &
\colhead{$\omega$}   &
\colhead{$\tau_{peri}$} &
\colhead{Epoch} &
\colhead{H$_v$} &
\colhead{Orbital Arc} &
\colhead{N$_{obs}$} &
\colhead{albedo$^b$} &
\colhead{Diameter} \\
\colhead{}   &
\colhead{}  &
\colhead{(AU)}  &
\colhead{}  &
\colhead{(deg)}   &
\colhead{(deg)}   &
\colhead{(deg)}   &
\colhead{(mjd)} &
\colhead{(mjd)} &
\colhead{(mag)} &
\colhead{(days)} &
\colhead{} &
\colhead{} &
\colhead{(km)} \\

}
\startdata
   1 &      39420 & 1.96 & 0.08 & 21.44 & 187.6 & 318.6 & 54364.16 & 53995 & 15.11 &  8 & 128 & 0.20 &  2.8 \\
   2 &  2006 RJ43 & 3.12 & 0.16 &  5.43 &  19.9 & 315.4 & 53851.05 & 53995 & 16.08 & 14 &  93 & 0.04 &  4.1 \\
   3 &  2006 ST62 & 3.08 & 0.17 &  7.63 & 187.8 & 160.2 & 53909.48 & 53996 & 16.22 & 14 &  89 & 0.04 &  3.8 \\
   4 &     145635 & 2.77 & 0.20 &  9.78 &  10.5 &  44.6 & 54140.97 & 53995 & 15.59 &  8 & 112 & 0.08 &  3.6 \\
   6 &      82495 & 2.70 & 0.08 &  5.48 &  19.0 & 353.0 & 54004.32 & 53995 & 16.05 &  8 &  77 & 0.08 &  2.9 \\
   7 & 2001 VZ123 & 3.03 & 0.16 &  6.10 &  15.7 &  21.2 & 54098.53 & 53996 & 16.21 &  8 &  64 & 0.04 &  3.8 \\
   8 &      70172 & 2.24 & 0.21 &  1.49 & 161.7 & 108.4 & 53740.54 & 53995 & 16.24 &  8 &  79 & 0.20 &  1.7 \\
   9 &  2006 RK43 & 2.76 & 0.12 &  8.94 &  14.7 & 237.2 & 53497.56 & 53995 & 16.16 & 14 &  85 & 0.08 &  2.8 \\
\enddata
\vskip 0.05in
\scriptsize{$^a$numbering and order are arbitrary and derive from the survey processing methods.  $^b$assumed from semimajor axis, see text. }
\label{tab.orb}
\end{deluxetable}

\begin{deluxetable}{crcccc}
\tablenum{3}
\tabletypesize{\scriptsize}
\tablecaption{Fitted Light Curve Parameters}
\tablewidth{0pt}
\tablehead{
\colhead{TALCS ID}   &
\colhead{MPC desig.}  &
\colhead{period}  &
\colhead{amplitude}  &
\colhead{$g'-r'$ color}   &
\colhead{U$^a$} \\
\colhead{}   &
\colhead{}  &
\colhead{(hr)}  &
\colhead{(mag)}  &
\colhead{(mag)}   &
\colhead{} 
}
\startdata

   1 &      39420 & $   105. \pm    21.$ & 1.18 & $1.913 \pm 0.50$ & 2 \\
   2 &  2006 RJ43 &     \dots            & \dots & $0.576 \pm 0.05$ & 0 \\
   3 &  2006 ST62 &     \dots            & \dots & $0.462 \pm 0.02$ & 0 \\
   4 &     145635 & $  6.905 \pm  0.002$ & 0.30 & $0.556 \pm 0.03$ & 3 \\
   6 &      82495 & $  10.95 \pm   0.02$ & 0.16 & $0.607 \pm 0.02$ & 2 \\
   7 & 2001 VZ123 & $  19.46 \pm   0.03$ & 0.85 & $0.329 \pm 0.03$ & 3 \\
   8 &      70172 & $ 5.7394 \pm 0.0004$ & 1.02 & $0.588 \pm 0.01$ & 3 \\
   9 &  2006 RK43 &     \dots            & \dots & $0.539 \pm 0.04$ & 0 \\
\enddata
\vskip 0.05in
\scriptsize{$^a$Light-curve fit reliability code as defined by \citet{harris83}; see text for details.  $^b$Detailed analysis of these objects indicates that their reliability code is more likely 1 than the initially determined 2; see text for details.  A value of ``\dots'' for period or color indicate insufficient data to make a measurement.}
\label{tab.lcfits}
\end{deluxetable}

\begin{deluxetable}{crcccc}
\tablenum{4}
\tabletypesize{\scriptsize}
\tablecaption{Followup Light Curve Periods}
\tablewidth{0pt}
\tablehead{
\colhead{TALCS ID}   &
\colhead{MPC desig.}  &
\colhead{initial period}  &
\colhead{U$^a$}   &
\colhead{new period} &
\colhead{amplitude} \\
\colhead{}  &
\colhead{}  &
\colhead{(hr)}  &
\colhead{}   &
\colhead{(hr)} &
\colhead{(mag)} 
}
\startdata

   1 &      39420 & $   105. \pm 21.$    & 2 & $>>70$ & $>0.5$ \\
  31 &      32705 & $    98. \pm 19.6$   & 2 & $>>70$ & $>0.5$ \\
  59 &  2002 UB14 & $ 2.7985 \pm 0.0005$ & 2 & $2.8\pm0.1$ & $0.2$ \\
 131 &      46748 & $  2.698 \pm 0.002$  & 2 & $2.7\pm0.1$ & $0.4$ \\
 248 &      73727 & $   17.5 \pm 3.$     & 2 & $14.2\pm0.1$ & $0.3$ \\
 250 &      24215 & $ 11.036 \pm 0.006$  & 2 & $11.2\pm0.1$ & $0.2$ \\
 254 &      12527 & $  7.613 \pm 0.004$  & 2 & \dots$~^b$ & $\ge0.3$ \\
 274 &     137987 & $    40. \pm 4.$     & 2 & \dots$~^c$ & \dots \\
 280 &      55523 & $    46. \pm 4.$     & 2 & \dots$~^c$ & \dots \\
 390 &      55924 & $   4.81 \pm 0.02$   & 3 & $4.8\pm0.1$ & $0.3$ \\

\hline
\enddata
\vskip 0.05in
\scriptsize{$^a$Light-curve fit reliability code as defined by \citet{harris83}; see text for details.\\$^b$Insufficient phase coverage to measure a full period.\\$^c$No observed variation above noise level.}
\label{tab.newobs}
\end{deluxetable}

\newpage

\clearpage

\newpage
\clearpage

\newpage
\clearpage

\newpage
\clearpage

\newpage
\clearpage

\newpage
\clearpage

\newpage
\clearpage

\newpage
\clearpage

\newpage
\clearpage

\newpage
\clearpage

\newpage
\clearpage

\newpage
\clearpage

\begin{thebibliography}{XXX}

\bibitem[Bertin \& Arnouts(1996)]{sex}
Bertin, E. \& Arnouts, S., 1996, ``SExtractor: Software for source extraction'', A\&A Supp., 117, 393.

\bibitem[Bottke et~al.(2002)]{bottkeAIII}
Bottke, W.F., Cellino, A., Paolicchi, P. \& Binzel, R.P., 2002, ``An Overview of the Asteroids:  The Asteroids III Perspective'', Asteroids III (W.~F.~Bottke, Jr. et~al., eds.), 3.


\bibitem[Bottke et~al.(2006)]{bottke06}
Bottke, W.F., Vokrouhlick\'{y}, D., Rubincam, D.P. \& Nesvorn\'{y}, D., 2006, ``The Yarkovsky and YORP Effects:  Implications for Asteroid Dynamics'', Annu. Rev. Earth Planet. Sci., 34, 157.

\bibitem[Boulade et~al.(2000)]{megacam}
Boulade, O., et~al., 2000, ``Development of MegaCam, the next-generation wide-field imaging camera for the 3.6-m Canada-France-Hawaii Telescope'', Proc. SPIE, 4008, 657.

\bibitem[\v{C}apek \& Vokrouhlick\'{y}(2004)]{capek04}
\v{C}apek, D. \& Vokrouhlick\'{y}, D., 2004, ``The YORP effect with finite thermal conductivity'', Icarus, 172, 526.

\bibitem[Chapman(2002)]{chapmanAIII}
Chapman, C.R., 2002, ``Cratering on Asteroids from {\it Galileo} and {\it NEAR Shoemaker}'', Asteroids III (W.~F.~Bottke, Jr. et~al., eds.), 315.

\bibitem[Dermawan(2004)]{dermawan}
Dermawan, B., 2004, ``Spin characteristics of very small main-belt asteroids'', Ph. D. thesis, School of Science, University of Tokyo.

\bibitem[\v{D}urech et~al.(2009)]{durech09}
\v{D}urech, J., Kaasalainen, M., Warner, B.D., Fauerbach, M., Marks, S.A., Fauvaud, S., Fauvaud, M., Vugnon, J.-M., Pilcher, F., Bernasconi, L. \& Behrend, R., 2009, ``Asteroid models from combined sparse and dense photometric data'', A\&A, 493, 291.

\bibitem[\v{D}urech et~al.(2007)]{durech07}
\v{D}urech, J., Scheirich, P., Kaasalainen, M., Grav, T., Jedicke, R. \& Denneau, L., 2007, ``Physical models of asteroids from sparse photometric data'', Proc. of IAU Symp. 236 (A. Milani, G.B. Valsecchi \& D. Vokrouhlick\'{y}, eds.), 191.

\bibitem[Gottleib(2001)]{gottleib}
Gottleib, A.D., 2001, ``Asymptotic accuracy of the jackknife variance estimator for certain smooth statistics'', arXiv:math/0109002.

\bibitem[Gywn(2008)]{megapipe}
Gwyn, S.D.J., 2008, ``MegaPipe: the MegaCam Image Stacking Pipeline at the Canadian Astronomical Data Centre'', PASP, 120, 212.

\bibitem[Harris(1996)]{harris96}
Harris, A.W., 1996, ``The Rotation Rates of Very Small Asteroids: Evidence for 'Rubble Pile' Structure'', Lunar and Planetary Sci., 27, 493.

\bibitem[Harris(2002)]{harris02}
Harris, A.W., 2002, ``On the Slow Rotation of Asteroids'', Icarus, 156, 184.

\bibitem[Harris \& Lagerros(2002)]{harrisAIII}
Harris, A.W. \& Lagerros, J.S.V., 2002, ``Asteroids in the Thermal Infrared'', Asteroids III (W.~F.~Bottke, Jr. et~al., eds.), 205.

\bibitem[Harris \& Young(1983)]{harris83}
Harris, A.W. \& Young, J.W., 1983, ``Asteroid Rotation: IV. 1979 Observations'', Icarus, 54, 59.

\bibitem[Harris et~al.(1989)]{harris89}
Harris, A.W., Young, J.W., Bowell, E., Martin, L.J., Millis, R.L., Poutanen, M., Scarltriti, F., Zappal\`{a}, V., Schober, H.J., Debenhogne, H. \& Zeigler, K.W., 1989, ``Photoelectric Observations of Asteroids 3, 24, 60, 261, and 863'', Icarus, 77, 171.

\bibitem[Henry et~al.(2008)]{henry08}
Henry, J.P., Evrard, A.E., Hoekstra, H., Babul, A. \& Mahdavi, A., 2008, ``The X-ray cluster normalization of the matter power spectrum'', arXiv:0809.3832.

\bibitem[Holsapple(2007)]{holsapple07}
Holsapple, K.A., 2007, ``Spin limits of Solar System bodies: From the small fast-rotators to 2003 EL61'', Icarus, 187, 500.

\bibitem[Ivezic et~al.(2001)]{ivezicSDSS}
Ivezic, Z., Tabachnik, S., Rafikov, R. et~al., 2001, ``Solar System Objects Observed in the Sloan Digital Sky Survey Commissioning Data'', AJ, 122, 2749.

\bibitem[Jedicke \& Herron(1997)]{jedicke97}
Jedicke, R. \& Herron, J.D., 1997, ``Observational Constraints on the Centaur Population'', Icarus, 127, 494.

\bibitem[Jedicke et~al.(2007)]{jedickePS}
Jedicke, R., Magnier, E.A., Kaiser, N. \& Chambers, K.C., 2007, ``The next decade of Solar System discovery with Pan-STARRS'', Proc. of IAU Symp. 236 (G.B. Valsecchi \& D. Vokrouhlick\'{y}, eds.), 341.

\bibitem[Kaasalainen(2004)]{mikko04}
Kaasalainen, M., 2004, ``Physical models of large number of asteroids from calibrated photometry sparse in time'', A\&A, 422, L39.

\bibitem[Kaasalainen \& \v{D}urech(2007)]{mikko07}
Kaasalainen, M. \& \v{D}urech, J., 2007, ``Inverse problems of NEO photometry:  Imaging the NEO population'', Proc. of IAU Symp. 236 (A. Milani, G.B. Valsecchi \& D. Vokrouhlick\'{y}, eds.), 151.

\bibitem[Kubica et~al.(2005)]{kubica}
Kubica, J., Denneau, L., Grav, T., Heasley, J., Jedicke, R., Masiero, J., Milani, A., Moore, A., Tholen, D. \& Wainscoat, R.~J., 2005, ``Efficient intra- and inter-night linking of asteroid detections using kd-trees'', Icarus, 189, 151.

\bibitem[Lacerda \& Jewitt(2007)]{pedro07}
Lacerda, P. \& Jewitt, D., 2007, ``Densities of Solar System objects from their rotational light curves'', AJ, 133, 1393.

\bibitem[Lacerda \& Luu(2003)]{pedro03}
Lacerda, P. \& Luu, J., 2003, ``On the detectability of lightcurves of Kuiper belt objects'', Icarus, 161, 174.

\bibitem[Lupton (1993)]{lupton93}
Lupton, R., 1993, ``Statistics in Theory and Practice'', Princeton University Press, 46.

\bibitem[Magnier \& Cuillandre(2004)]{elixir}
Megnier, E.A. \& Cuillandre, J.-C., 2004, ``The Elixir System: Data Characterization and Calibration at the Canada-France-Hawaii Telescope'', PASP, 116, 449.

\bibitem[Milani et~al.(2007)]{milani}
Milani, A., Gronchi, G.F., Farnocchia, D., Kne\u{z}evi\'{c}, Z.,
Jedicke, R., Denneau, L. \& Pierfederici, F., 2008, ``Topocentric
orbit determination: Algorithms for the next generation surveys'',
Icarus, 195, 474.

\bibitem[Muinonen(1996)]{mui96} 
Muinonen, K.\ 1996.\ Light Scattering by Gaussian Random Particles.\
Earth Moon and Planets 72, 339-342.

\bibitem[Muinonen(1998)]{mui98} 
Muinonen, K.\ 1998.\ Introducing the Gaussian shape hypothesis for
asteroids and comets.\ Astronomy and Astrophysics 332, 1087-1098.

\bibitem[Polishook \& Brosch(2008)]{polishook08}
Polishook, D. \& Brosch, N., 2008, ``Photometry and Spin Rate Distribution of Small-Sized Main Belt Asteroids'', arXiv:0811.1223.

\bibitem[Pravec \& Harris(2000)]{pravec00}
Pravec, P. \& Harris, A.W., 2000, ``Fast and Slow Rotation of Asteroids'', Icarus, 148, 12.

\bibitem[Pravec et~al.(2002)]{pravecAIII}
Pravec, P., Harris, A.~W., Micha\'{l}owski, T., 2002, ``Asteroid Rotations'', Asteroids III (W.~F.~Bottke, Jr. et~al., eds.), 113.

\bibitem[Pravec et~al.(2005)]{pravec05}
Pravec, P., Harris, A.W., Scheirich, P., et~al., 2005, ``Tumbling asteroids'', Icarus, 173, 108.

\bibitem[Pravec et~al.(2008)]{pravec08}
Pravec, P., Harris, A.W., Vokrouhlick\'{y}, D., et~al., 2008, ``Spin rate distribution of small asteroids'', Icarus, 197, 497.

\bibitem[Pope \& Szapudi(2008)]{pope08}
Pope, A.C. \& Szapudi, I., 2008, ``Shrinkage Estimation of the Power Spectrum Covariance Matrix'', MNRAS, 389, 766.

\bibitem[Rossi, Marzari \& Scheeres(2009)]{rossi09}
Rossi, A., Marzari, F. \& Scheeres, D.J., ``Computing the effects of YORP on the spin rate distribution of the NEO population'', in prep.

\bibitem[Rubincam(2000)]{rubincam00}
Rubincam, D.P., 2000, ``Radiative Spin-up and Spin-down of Small Asteroids'', Icarus, 148, 2.

\bibitem[Salo(1987)]{salo87}
Salo, H., 1987, ``Numerical Simulations of Collisions between Rotating Particles'', Icarus, 70, 37.

\bibitem[Scheeres(2007a)]{scheeres07yorp}
Scheeres, D., 2007, ``The dynamical evolution of uniformly rotating asteroids subject to YORP'', Icarus, 188, 430.

\bibitem[Scheeres(2007b)]{scheeres07bin}
Scheeres, D., 2007, ``Rotational fission of contact binary asteroids'', Icarus, 189, 370.

\bibitem[Slivan et~al.(2003)]{slivan03}
Slivan, S.M., Binzel, R.P., Crespo da Silva, L.D., Kaasalainen, M., Lyndaker, M.M., \& Kr\u{c}o, M., 2003, ``Spin vectors in the Koronis family: comprehensive results from two independent analyses of 213 rotation lightcurves'', Icarus, 162, 285.


\bibitem[Sullivan et~al.(2002)]{sullivanAIII}
Sullivan, R.J., Thomas, P.C., Murchie, S.L. \& Robinson, M.S., 2002, ``Asteroid Geology from {\it Galileo} and {\it NEAR Shoemaker} Data'', Asteroids III (W.~F.~Bottke, Jr. et~al., eds.), 331.

\bibitem[Tedesco et~al.(2005)]{tedesco05}
Tedesco, E.F., Cellino, A., Zappal\`{a}, V., 2005, ``The Statistical Asteroid Model. I. The Main-Belt Population for Diameters Greater than 1 Kilometer'', AJ, 129, 2869.

\bibitem[Tedesco et~al.(2002)]{tedesco02}
Tedesco, E.F., Noah, P.V., Noah, M. \& Price, S.D., 2002, ``The Supplemental IRAS Minor Planet Survey'', AJ, 123, 1056.


\bibitem[Vokrouhlick\'{y} et~al.(2007)]{vok07}
Vokrouhlick\'{y}, D., Breiter, S., Nesvorn\'{y}, D. \& Bottke, W.F., 2007, ``Generalized YORP evolution: Onset of tumbling and new asymptotic states'', Icarus, 191, 636.

\bibitem[Vokrouhlick\'{y} \& \v{C}apek(2002)]{vok02}
Vokrouhlick\'{y}, D. \& \v{C}apek, D., 2002, ``YORP-Induced Long-Term Evolution of the Spin State of Small Asteroids and Meteoroids:  Rubincam's Approximation'', Icarus, 159, 449.

\bibitem[Walsh et~al.(2008)]{walsh08}
Walsh, K.J., Richardson, D.C. \& Michel, P., 2008, ``Rotational breakup as the origin of small binary asteroids'', Nature, 454, 188.

\end{thebibliography}
\end{document}